\documentclass[preprint,10pt]{elsarticle}

\usepackage{amssymb}
\usepackage{fancyheadings}
\usepackage{amsmath,mathrsfs,mathtools,mathdots,amssymb,commath,bm}
\usepackage{subfigure}
\usepackage{soul}
\usepackage{chemarrow}
\usepackage{nomencl}
\usepackage{multirow,makecell}
\usepackage[numbers]{natbib}
\usepackage{booktabs}
\usepackage{wasysym} 
\usepackage{graphicx}
\usepackage{array}
\usepackage{float}
\usepackage{multirow}

\usepackage{hyperref}
\hypersetup{
    colorlinks=true,
    linkcolor=blue,
    citecolor=blue,
    urlcolor=blue
}

\journal{Journal Computational Physics}

\begin{document}

\begin{frontmatter}

\title{Extension of the consistent $\delta^{+}$-SPH model for multiphase flows considering the compressibility of different phases}

\author[label1]{Xiao-Ting Huang}
\author[label1]{Peng-Nan Sun}
	\corref{cor1}\ead{sunpn@mail.sysu.edu.cn}
\author[label1]{Hong-Guan Lyu}
\author[label2]{Andrea Colagrossi}
\author[label3]{A-Man Zhang}
	\address[label1]{School of Ocean Engineering and Technology, Sun Yat-sen University, Zhuhai, China}
\address[label2]{Institute of Marine Engineering (INM), National Research Council (CNR), Rome, Italy}
\address[label3]{College of Shipbuilding Engineering, Harbin Engineering University, Harbin, China}


\begin{abstract}
  In hydrodynamic problems involving wave impact on structures, air compressibility is crucial for accurate pressure prediction when an air bubble is entrapped. In this work, the consistent $\delta^{+}$-SPH model, originally developed for single-phase scenarios, is extended to multiphase contexts.
  Although the consistent $\delta^{+}$-SPH model shows good performance for single phase and viscous flow simulations, extending it to multiphase scenarios presents challenges, such as proper implementation of particle shifting for multiphase interfaces. 
  Therefore, within the framework of the consistent $\delta^{+}$-SPH, we introduce the following  enhancements: 
  firstly, new strategy for handling $\delta \bm{u}$-terms given by the particle shifting technique at multiphase interfaces are proposed to maintain stability and conservation.  
  Secondly, for modeling of incompressible phases, like water, an acoustic damper term is introduced to alleviate acoustic waves resulting from the weakly-compressible assumption, which is expected to achieve smooth pressure field comparable to truly-incompressible hypothesis, thereby reducing the nonphysical pressure wave during the violent impact state; for modeling compressible phases like air, a physical sound speed is adopted in the equation of state to accurately model real gas phase compressibility.
  To test and validate the present multiphase SPH model, simulations were conducted for six scenarios.
  In particular, except for sloshing with two-layer liquids, the other scenarios fully consider air pressure oscillations when air is entrapped, compressed, or expanded by surrounding flows. 
  The results demonstrate significant advantages of the present SPH model in simulating multiphase problems involving strong liquid impact and different phase compressibility.
\end{abstract}


\end{frontmatter}
\section{Introduction}
The study of multiphase flows is essential in both scientific and engineering domains, attracting significant attention from scholars in recent years \cite{guan2022research,prasad2024comprehensive}.
Computational Fluid Dynamics (CFD) has emerged as a valuable approach for studying multiphase flow problems. However, due to the unavoidable discontinuous characteristics such as density, and physical interactions at the phase interfaces, Eulerian mesh-based methods face inherent challenges.
In this regard, Lagrangian particle methods stand out for their ability to handle complex interfaces and multiphase flows with deformations without requiring additional numerical treatments for interface tracking \cite{monaghan1995sph,wen2021multiphase,duan2019novel,han2024incompressible}. Among these methods, Smoothed Particle Hydrodynamics (SPH) is extensively applied in simulating multiphase flows \cite{hu2006multi,szewc2012study,meng2020multiphase,luo2021particle,he2023numerical,zhao2023multi,pozorski2024smoothed}. 

Previous studies on multiphase flow models, such as water-gas phase models, often assume that both water and gas are incompressible, as seen in VOF-based methods (see e.g.,  \cite{zhao2010numerical}) and level-set methods (see e.g., \cite{kleefsman2005volume}). However, in scenarios involving violent free surface flows, such as wave impacts on structures or sloshing, air entrainment or water-air mixing can lead to compression and expansion of the air phase, resulting in pressure changing rapidly and with long-term oscillation \cite{bagnold1939interim,takahashi1985uplift,lyu2023establishment,lyu2025establishment-v2}. Hence, it is crucial to consider the compressibility of different phases when studying such multiphase flow problems.

Multiphase SPH schemes can be categorized according to how pressure is solved, leading to two main approaches: Weakly Compressible SPH (WCSPH) \cite{dalrymple2006numerical,manenti2018standard} and Incompressible SPH (ISPH) \cite{hu2007incompressible,rezavand2018isph,gotoh2014enhancement}. 
In the WCSPH framework, the pressure is explicitly solved based on the Equation of State (EOS), making it well suited for compressible multiphase flow problems, while the ISPH model solves the pressure implicitly via the Poisson equation, which poses challenges for handling strongly compressible flows \cite{liu2003smoothed,LIND2016129, le2025smoothed}. Additionally, the ISPH scheme also faces difficulties in parallelization \cite{o2022eulerian}, while the WCSPH scheme is well suited for multi-GPU simulations \cite{lyu2023sphydro}. In this study, the WCSPH scheme is chosen because of its ability to handle the compressibility of fluid phases while naturally satisfying the free surface condition \cite{colagrossi2009theoretical}.

Within the context of WCSPH multiphase flows, two main challenges we discuss in the present work are: (i) numerical instability, such as pressure instability, particle penetration at phase interfaces; and (ii) modeling the compressibility of different fluid phases. Various approaches have been proposed to tackle the two challenges, which are discussed below. 

Pressure instability at multiphase interfaces is often attributed to density discontinuity. To address this issue, particular attention is paid to the treatment of density. 
Hu et al. \cite{HU2006844} used the summation of density to obtain smoothed pressure field for multiphase SPH scheme, but under the assumption of uniform particle distribution. Furthermore, it does not work for free-surface problems due to the truncation of the kernel support. Another approach for density evaluation is using the continuity equation, which has been widely used and well-supported by numerical techniques. For example, Colagrossi and Landrini \cite{colagrossi2003numerical} used moving-least-squares (MLS) interpolation to frequently reinitialize density at the air-water interfaces, but this introduced numerical instabilities near the free surface.
Hammani et al. \cite{hammani2020detailed} extended the $\delta$-SPH model to multiphase flow, introducing a diffusive term to improve the pressure field.
Recently, Guo et al. \cite{GUO2024113336} proposed a new discretization scheme for the divergence of velocity to ensure the stability of interfaces, indicating that the SPH operators in the continuity equation at multiphase interfaces should be carefully conducted.  
 
Non-uniform particle distribution is another cause of instability at interfaces. A well-acknowledged numerical technique is Particle Shifting Technique (PST), which shows a strong capability to maintain uniform distribution \cite{xu2009accuracy,lind2012incompressible,mokos2017multi,khayyer2019projection,rezavand2020weakly} for multiphase flows problems. Specially, Oger et al. \cite{OGER201676} adjusted particle trajectories using an Arbitrary Lagrangian-Eulerian (ALE) formalism by applying a small velocity perturbation to correct Lagrangian velocity, which is shown to be accurate and stable for free-surface flows. 
More recently, ALE frameworks for the consistent $\delta^{+}$-SPH \cite{sun2019consistent} and the $\delta$-ALE-SPH \cite{antuono2021delta} model, have been proposed as good alternative approaches to improve volume conservation. However, these methods are primarily employed for single-phase flows. Further, when involving multiphase scenarios, the PST at the multiphase interfaces needs close attention to keep numerical stability and a proper implementation of PST for multiphase flows is still an open problem. 

Moreover, applying background pressure is another effective method for stabilizing interfaces by preventing particle clumping due to negative pressure \cite{chen2015sph,zhu2018improved,he2022stable}. However, the value of background pressure for phases should be carefully selected, \textcolor{black}{since excessive background pressure can lead to non-physical dissipation.} 
\textcolor{black}{In addition, Grenier et al. ~\cite{grenier2009hamiltonian} introduced an artificial repulsive force to avoid the inter-penetration phenomenon. Similarly, Monaghan and Rafiee \cite{monaghan2013simple} proposed a repulsive force related to the ratio of the densities of the fluids. Recently, Guo et al. \cite{GUO2024113336} introduced an improved interface force based on the work of Monaghan and Rafiee \cite{monaghan2013simple} to better preserve the sharpness of the interface in violent flows. These approaches have been widely adopted by many studies to maintain the sharpness of the interface \cite{lyu2021study,FANG2022110789,huang2022water,ju2023study}. Nevertheless, it often requires choosing parameters carefully to avoid introducing excessive numerical disruption. In some cases, it may lead to a gap between different fluid phases.}

When considering the modeling of phase compressibility, the SPH method primarily focuses on strongly compressible problems such as bubble merging \cite{sun2021accurate,sun2021accurateb}. In these scenarios, the compressibility of the bubble is typically modeled using various Tait equations. Although the effect of air in situations like dam breaking \cite{guilcher2014simulation,YOO2024112930}, the flooding of a damaged cabin \cite{CAO20187}, and slamming of the flat plate \cite{FANG2022110789} has been explored using multiphase SPH models, the discussion on the stage that an air cavity becomes closed or entrapped, is still insufficient. This is mainly due to the difficulties in accurately predicting the oscillation of air pressure caused by compressibility in current multiphase SPH schemes. As aforementioned, it is still necessary to develop an accurate and stable multiphase SPH model that is capable of high stability free from the use of background pressure, repulsive force methods and considering the real compressibility of phases. 

Due to the good performance of the consistent $\delta^{+}$-SPH \cite{sun2019consistent} to simulate violent flows, we extend it to multiphase problems and model the compressibility of air using the real physical sound speed. With the new numerical treatment of shifting velocity at the multiphase interfaces, good stability at the interfaces is achieved without the need for additional background pressure. In the present work, six cases are simulated, including a two-phase hydrostatic test, \textcolor{black}{slamming of  insulation panel in the Liquefied Natural Gas (LNG) tank}, water entry of wedge, two-layer liquid sloshing, and sloshing with entrapped air. The ability of present multiphase $\delta^{+}$-SPH to consider flow compressibility accurately is highlighted. 

The paper is arranged as follows: Section \ref{numerical-model} details the extended consistent $\delta^{+}$-SPH model for multiphase flow problems, and the test cases are presented and discussed in Section \ref{section:results}. The conclusion is given in Section \ref{conclusion}. \textcolor{black}{In addition, the tests of strategies for the treatment of \texorpdfstring{$\delta \bm{u}$}--terms are discussed in \ref{section:appendix}.}
\section{Establishment of the multiphase consistent $\delta^{+}$-SPH model} \label{numerical-model}
\subsection{Governing equation}
In the pure Lagrangian framework, the general governing equation for weakly compressible fluid is written as:
\begin{equation} \label{eq:N-s-equation}
\begin{cases}
\frac{\Dif \rho}{\Dif t} &= -\rho\nabla\cdot\boldsymbol{u},\\
\rho\frac{\Dif\boldsymbol{u}}{\Dif t}&= -\nabla p+ \nabla\cdot\mathbb{V} + \rho\bm{g},\\
\frac{\Dif \boldsymbol{r}}{\Dif t} &= \boldsymbol{u} ,\,\,\,\,\,\,\,\, p = F(\rho), 
 \end{cases}
\end{equation}
where $\rho$, $\boldsymbol{u}$, $\boldsymbol{r}$ and $p$ represent density, velocity, position and pressure, respectively. $\boldsymbol{g}$ represents a generic volume force. $F$ relates the pressure field to the density. Specifically, in the present weakly compressible framework, pressure is explicitly determined using the Tait equation:
\begin{equation}\label{eq:eos}
  p=B\left[\left(\frac{\rho}{\rho_0}\right)^{\gamma}-1\right]+p_b,\quad  B=\frac{\rho_{0}c_{0}^2}{\gamma}.
\end{equation}

In Eq. \ref{eq:eos}, $\rho_0$ represents the reference density of the fluid and $\gamma$ is the adiabatic coefficient, typically set to $\gamma_w =7.0$ for the liquid phase and $\gamma_a =1.4$ for gas. These values are widely used in water-gas multiphase studies \cite{hao2004numerical}. Here, $p_b$ denotes the background pressure. In several previous SPH studies on multiphase flows, a uniform pressure $p_b$ has been assumed to facilitate non-penetration at interfaces \cite{chen2015sph,zhu2018improved, hammani2020detailed}. However, this approach has to determine the numerical value of background pressure suitably as excessive value may cause non-physical dissipation. 
To avoid this, $p_b =0$ is set both for the gas and water phase and stable pressure field is obtained as well without relying on background pressure in the present scheme. 

The sound speed, $c_0$, accounts for the compressibility of different phases. For liquid phase like water, \textcolor{black}{the sound speed $c_{0w}$ is based on the weakly compressible assumption, $c_{0w} \geq 10\max \left( U_{\max},\sqrt{{p_{\max}}/\rho_{0w}}\right)$ where $U_{\max}$ is the maximum expected fluid velocity, $p_{\max}$ is the maximum expected fluid pressure, and $\rho_{0w}$ is the reference density of water.}

\textcolor{black}{Notably, recent novel multiphase schemes allow reducing sound speeds for gas phase (e.g. \cite{chen2015sph,meng2020multiphase,hammani2020detailed}), instead of relying solely on real physical sound speeds, to improve numerical efficiency.}
However, in this study, to model real compressibility accurately when gas is entrapped, the numerical sound speed of the gas phase is set to the real physical sound speed, i.e., $c_{0a} = 340$ m/s.

\subsection{Consistent $\delta^{+}$-SPH model}
\subsubsection{Discretized governing equation}
The shifting velocity $\delta \bm{u}$ is straightforwardly included in the governing equation reformulated into a quasi-Lagrangian framework \textcolor{black}{following the studies \cite{ADAMI2013292,LIND2016129,zhang2017generalized,sun2019consistent,antuono2021delta}}; the quasi-Lagrangian derivative is given by:
\begin{equation}\label{eq:ale-time-derivative}
  \frac{\dif f}{\dif t}=\frac{\Dif f}{\Dif t}+\delta\boldsymbol{u}\cdot\nabla f. 
\end{equation} 
\textcolor{black}{The way to calculate the shifting velocity $\delta \bm{u}$ in our framework will be discussed further down, in Section \ref{sec:shifting_velocity}.} When $\delta \bm{u} = 0$, the framework becomes a pure Lagrangian formalism. 

\textcolor{black}{Utilizing the definition provided in Eq. \ref{eq:ale-time-derivative},} the consistent $\delta^{+}$-SPH multiphase model within a quasi-Lagrangian framework \cite{sun2019consistent} is following:
\begin{equation}\label{eq:ale-governing}
  \begin{cases}
    \dfrac{\dif\rho}{\dif t}&=-\rho\nabla \cdot\boldsymbol{u} -\rho\nabla\cdot\delta\boldsymbol{u} + \nabla \cdot(\rho\delta\boldsymbol{u}),\\
    \rho\dfrac{\dif\boldsymbol{u}}{\dif t}&=-{\nabla p}+\nabla \cdot\mathbb{V}+\rho\boldsymbol{g}+\nabla \cdot(\rho\boldsymbol{u}\otimes\delta\boldsymbol{u}) \textcolor{black}{-\rho\bm{u}\nabla\cdot\left(\delta\boldsymbol{u}\right)},\\
    \dfrac{\dif \boldsymbol{r}}{\dif t}&=\boldsymbol{u}+\delta\boldsymbol{u},\quad p=F(\rho).
  \end{cases}
\end{equation}
The system in Eq. \ref{eq:ale-governing} is identical to the single-phase consistent $\delta^{+}$-SPH model. However, aside from the treatment of the equation of state, this paper shows that the divergence terms given by $\delta \bm{u}$ are crucial for modeling multiphase consistent $\delta^{+}$-SPH, especially in the continuity equation. \textcolor{black}{Notably, the last term $\rho\bm{u}\nabla\cdot\left(\delta\boldsymbol{u}\right)$ within the momentum equation of the Eq. \ref{eq:ale-governing} is not discretized in the following work. This simplification has also been used in the previous works \cite{zhang2017generalized,sun2019consistent,michel2023energy,MARRONE2025117473}. As demonstrated by Sun et al. \cite{sun2019consistent}, including or not this term in the scheme does not seem to induce sensible differences in the final results.} 

Using the SPH approximation (details in \cite{liu2010smoothed,monaghan1994simulating}), the Eq. \ref{eq:ale-governing} is discretized, and \textcolor{black}{four key techniques are used to further improve numerical accuracy, including:}
 \begin{enumerate} 
  \item Using the Tensile Instability Control (TIC) \cite{sun2018multi} technique to avoid numerical voids. 
  \item Adding an acoustic damper term in the momentum equation for the incompressible phase to achieve a smooth pressure field comparable to that in the truly incompressible hypothesis. 
  \item Using PST to obtain uniform particle distribution and integrating advection velocity $\bm{u} + \delta \bm{u}$ into the governing equations for the multiphase ${\delta^{+}}$-SPH scheme. 
  \item Introducing numerical treatments of the shifting velocity at interfaces, including fluid-solid and multiphase interfaces, to maintain a stable pressure field, as described in Section \ref{section:interface-treatment}.
\end{enumerate}
Considering these points, the present SPH scheme for multiphase flows based on the consistent $\delta^{+}$-SPH is discretized as follows:
 \begin{equation}\label{eq:final-ale-equation} 
   \begin{aligned}
 &\begin{cases}
    \dfrac{\dif \rho_i}{\dif t}&=-\rho_i\langle\nabla \cdot\boldsymbol{u}\rangle_i
    -\rho_i \underbrace{\langle\nabla \cdot\delta\boldsymbol{u}\rangle_i}_{\text{$\delta \bm{u}$-term}}
    +\underbrace{\langle\nabla \cdot(\rho\delta\boldsymbol{u})\rangle_i}_{\text{$\delta \bm{u}$-term}} +\delta h c_0\mathcal{D}_i,
    \\
    {\rho_i}\dfrac{\dif \boldsymbol{u}_i}{\dif t}&= -{\langle\nabla p\rangle_i } +\langle\nabla \cdot\mathbb{V}\rangle_i+{\rho_i}\boldsymbol{g}+ \underbrace{\langle\nabla \cdot({\rho}\boldsymbol{u}\otimes\delta\boldsymbol{u})\rangle_i}_{\text{$\delta \bm{u}$-term}} + \boldsymbol{F}^{ad}_{i},
    \\
    \dfrac{\dif \boldsymbol{r}_i}{\dif t}&=\boldsymbol{u}_i+\delta\boldsymbol{u}_i,\quad p_i=F(\rho_i),
\end{cases}\\
   &\text{with} \\
  &\boldsymbol{F}^{ad}_{i} =
  \begin{cases}
  \alpha_2 \rho_i c_0 h \langle\nabla \left(\nabla \cdot \boldsymbol{u}\left(\boldsymbol{r}\right)\right)\rangle_i \,, &  i \in \Omega_f^{I}  \\
  0 \,, & \text{otherwise},
\end{cases}
\end{aligned}
 \end{equation}
where the subscript $j$ represents the neighboring particles of particle $i$. The diffusive coefficient $\delta = 0.1$ is used. The term $\mathcal{D}_i$ denotes the numerical diffusive term proposed by \cite{antuono2012numerical}, aimed at filtering out high-frequency pressure vibrations:
\begin{equation}\label{d}
  \textcolor{black}{
  \mathcal{D}_i=\sum\limits_{j,\atop  \Omega(j) = \Omega(i)} \psi_{ji}\frac{\boldsymbol{r}_{ji}\cdot\nabla_iW_{ij}}{\left\|\boldsymbol{r}_{ji}\right\|^2}V_j, \quad \psi_{ij} = 2\left(\rho_j-\rho_i\right)-\left[\langle\nabla\rho\rangle_i^L+\langle\nabla\rho\rangle_j^L\right]\cdot\boldsymbol{r}_{ji},
  }
\end{equation} 
where $\boldsymbol{r}_{ji} = \boldsymbol{r}_{j} - \boldsymbol{r}_{i}$, and the renormalized density gradient $\langle\nabla\rho\rangle^L$ is defined as:
\begin{equation}\label{gradient-density}
  \textcolor{black}{
  \langle\nabla\rho\rangle_{i}^{L}=\sum\limits_{j,\atop  \Omega(j) = \Omega(i)}(\rho_j-\rho_i)\mathbb{L}_i\nabla W_{ij}V_j,\quad\mathbb{L}_i =\left[\sum\limits_{j,\atop  \Omega(j) = \Omega(i)}\boldsymbol{r}_{ji}\otimes\nabla W_{ij}V_j\right]^{-1}.
  }
\end{equation}
\textcolor{black}{To prevent non-physical density diffusion between different fluids, the numerical diffusive term is applied only for particles within the same fluid set, i.e. $\Omega(j) = \Omega(i)$ (see Figure \ref{fig:particle-set}), following the work by Hammani et al. \cite{hammani2020detailed}; it means that the summations in Eqs. \ref{d} and \ref{gradient-density} all apply to particles belonging to the $i$th particle set $\Omega(i)$.}

\textcolor{black}{$W_{ij} = W\left(r_{ji}, h\right)$ represents the kernel function where $r_{ji} = \|\bm{r}_{j} - \bm{r}_i\|$, and $h$ denotes the smoothing length. 
In the present work, a Gaussian kernel function \cite{grenier2009hamiltonian} with compact support is employed:}
\begin{equation}
  \textcolor{black}{
  W (r,h) =
  \begin{cases} 
    \frac{1} {\pi h^{2}} \left[ \frac{e^{-( r / h )^{2}}-C_{0}} {1-C_{1}} \right], & \text{if } r \leqslant\epsilon h \\  
    0, & \text{otherwise}
  \end{cases},  \quad
  C_{0}=e^{-\epsilon^{2}}, \quad C_{1}=C_{0} ( 1+\epsilon^{2} ).}
  \end{equation}
  \textcolor{black}{The length $\epsilon h $ represents a cut-off radius, set equal to $3h$, and $h/\Delta x = 1.33$ where $\Delta x$ is averaged particle spacing at initialization.}

Furthermore, the acoustic damper term $\boldsymbol{F}^{ad}_{i}$ is incorporated for incompressible fluids (referred to as $\Omega_f^{I}$) to minimize the effects associated with the weakly compressible assumption, as demonstrated in \cite{sun2023inclusion}. The parameter $\alpha_2$ needs to be large enough to dampen acoustic waves generated during violent impact events. However, an excessively large $\alpha_2$ can significantly reduce the time step (details in \cite{sun2023inclusion}). In multiphase scenarios, since the time step is constrained by the physical gas sound speed (see Subsection \ref{subsubsection:timestep}), which is sufficiently small, thus the parameter $\alpha_2$ can be set to $\alpha_2=10$. This ensures smoother pressure field for the incompressible phase, closely resembling predictions under the incompressible hypothesis.

\subsubsection{Discretized SPH differential operators}
\begin{figure}
  \centering
  \includegraphics[width=0.85\linewidth]{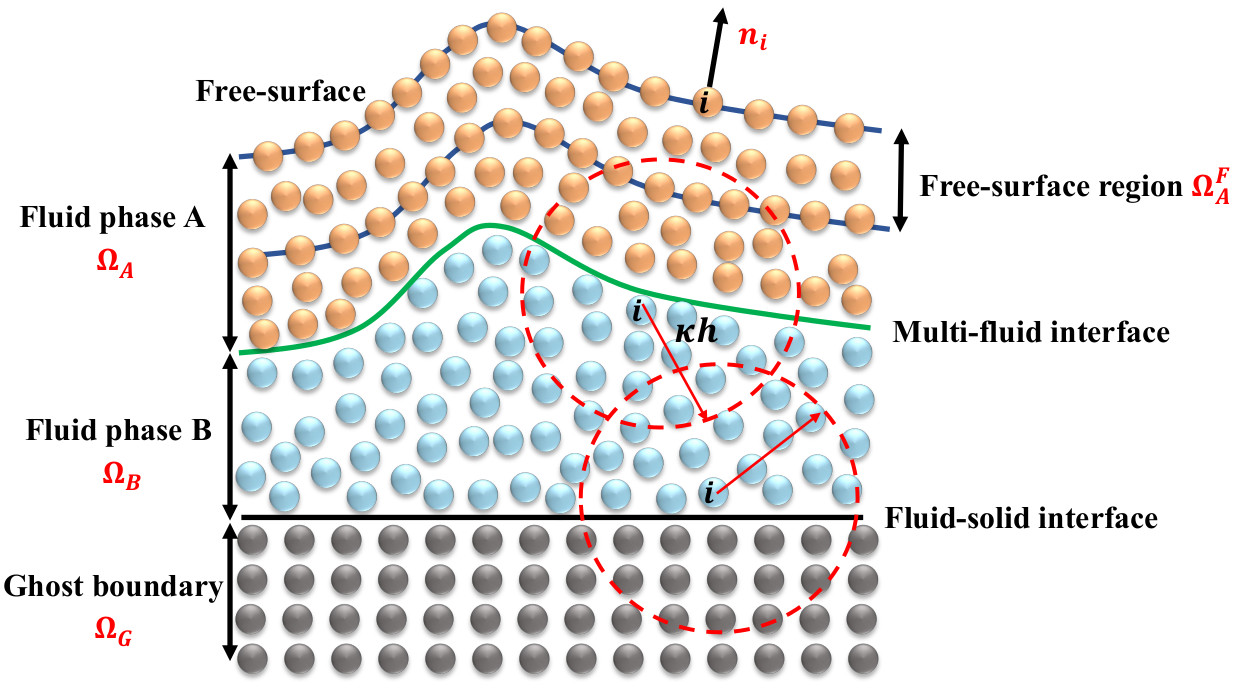}
  \caption{Schematic diagrams with particle set definitions, where $\Omega$ represents the universe of SPH particles and $\Omega = \Omega_{A} \cup \Omega_{B} \cup \Omega_{G}$. $\Omega_{A}^{F}$refers particles that belong to $\Omega_{A}$ and are free-surfaces particles.} \label{fig:particle-set}
\end{figure}

Now, the discretization of the differential operators on the right side of Eq. \ref{eq:final-ale-equation} is introduced. These differential operators are approximated through the convolution with the kernel function. Based on the classic discretization of the pressure gradient, the technique of TIC \cite{sun2018multi} is further employed to avoid numerical voids. Specifically, when the pressure $p_i$ of particle $i$ located outside the free surface region $\Omega_A^{F}$ (see Figure \ref{fig:particle-set}) is negative, the classic term $p_i + p_j$ is replaced by $p_j - p_i$. 
Conversely, particle $i$ with positive pressure, or located in the free surface region, is still calculated in the classic form. \textcolor{black}{The detection of the free surface region $\Omega_A^{F}$ is conducted following the procedure established by \cite{marrone2010fast}.}
Therefore, the pressure gradient $\nabla p$ is discretized as:
\begin{equation}
  \begin{cases}
    \langle\nabla p\rangle_{i}=\sum\limits_{j}\left(p_{j}+p_{i}\right)\nabla_{i}W_{ij}V_{j} + S_i \sum\limits_{j}\nabla_{i}W_{ij}V_{j},\\
    S_i=
    \begin{cases}
      2p_i&\quad\mathrm{if~}p_i<0\mathrm{~and~}i\notin\Omega_{A}^{F},\\
      0&\quad\mathrm{otherwise~}.
    \end{cases}
  \end{cases} 
\end{equation}   

The divergence of the velocity $\nabla \cdot \bm{u}$ and  viscous stress tensor $\nabla \cdot\mathbb{\mathbb{V}}$ are approximated through below formulas (see \cite{colagrossi2009theoretical, colagrossi2011theoretical} for more details):  
\begin{equation}\label{eq:divergence-velocity}
\begin{cases}
  \langle\nabla \cdot\bm{u}\rangle_{i}= \sum\limits_{j}\left(\bm{u}_j-\bm{u}_i\right)\cdot\nabla_{i}W_{ij}V_{j},\\
  \langle\nabla \cdot\mathbb{\mathbb{V} }\rangle_{i}= K\mu\sum\limits_{j}\pi_{ij}\nabla_{i}W_{ij}V_{j},\\
  \pi_{ij}=\frac{\bm{u}_{ji}\cdot \bm{r}_{ji}}{||\bm{r}_{ji}||^2},
\end{cases}
\end{equation}
where ${\mu}$ denotes the dynamic viscosity of the fluid; parameter $K = 2(n + 2)$ with $n$ the dimension of the simulations. In cases involving violent impact flows, the coefficient $K\mu$ is directly replaced by $\alpha c_0 \rho_0 h$, $\alpha$ referred to as the artificial viscosity parameter, aiming to improve numerical stability. More details can be found in \cite{antuono2012numerical}. The coefficient $\alpha$ is generally recommended to vary between 0 and 0.1 for violent impact problems, so a value of $\alpha = 0.1$ is adopted in the present work, unless otherwise specified. 

Following \cite{antuono2021delta,sun2019consistent}, the approximation of the terms given by shifting velocity $\delta \bm{u}$ (called $\delta \bm{u}$-terms hereinafter) is written as:
\begin{equation} \label{eq:delta-uterms}
  \fbox{\(\delta \bm{u}\)-terms}
\begin{dcases}
  \left\langle\nabla \cdot(\delta\boldsymbol{u})\right\rangle_i=\sum\limits_{j}(\delta\boldsymbol{u}_j-\delta\boldsymbol{u}_i)\cdot{\nabla}_iW_{ij}V_j,\\
  \left\langle\nabla \cdot(\rho\delta\boldsymbol{u})\right\rangle_i=\sum\limits_{j}(\rho_i\delta\boldsymbol{u}_i+\rho_j\delta\boldsymbol{u}_j)\cdot{\nabla}_iW_{ij}V_j,\\
  \left\langle\nabla \cdot(\rho\boldsymbol{u}\otimes\delta\boldsymbol{u}) \right\rangle_i=  \sum\limits_{j}(\rho_i\boldsymbol{u}_i\otimes\delta\boldsymbol{u}_i+\rho_j\boldsymbol{u}_j\otimes\delta\boldsymbol{u}_j)\cdot{\nabla}_iW_{ij}V_j.
\end{dcases}
\end{equation}
\textcolor{black}{Specifically, the first $\delta \boldsymbol{u}$-term $\nabla \cdot(\delta\boldsymbol{u})$ follows the same discretization strategy as the velocity divergence in Eq. \ref{eq:divergence-velocity}, using $(f_j - f_i)$ differential terms where $f$ represents a vector variable. The other two divergence operators adopt symmetric $(f_i + f_j)$ summation terms. Owing to this symmetric structure, the conservation of mass and linear momentum can be maintained \cite{antuono2021delta}.}

\subsubsection{Boundary implementations and time step}\label{subsubsection:timestep}
For boundary conditions, mirror ghost particle techniques theoretically provide superior accuracy and convergence, while the fixed ghost technique introduces minor numerical discrepancies in the SPH scheme. However, the practical applicability of the former is limited in engineering contexts, such as water entry simulations. Therefore, the fixed ghost particle technique is employed, as the numerical errors in the cases examined in this paper are negligible.
Additionally, for potential particle penetration issues in particle-based boundaries, a strategy proposed in \cite{LYU2024106144}, which involves correcting the normal velocity of potentially penetrating particles near ghost particles, can be a viable alternative. 

The discrete governing equation is updated in time using the fourth-order Rang-Kutta time integration.  
The time step $\Delta t$  has to account for maximum acceleration $|\bm{a}_\text{max}|$ in the flows, the sound speed, viscous diffusion.  It is determined based on the following conditions \cite{sun2023inclusion}:
\begin{equation}\label{eq:timestep}  
  \begin{cases}
  \Delta t_{a} = 0.25\sqrt{\frac{h}{|\bm{a}_{\text{max}}|}},\\
    \Delta t_{c} = \text{CFL min}(\frac{h}{c_{0w}},  \frac{h}{c_{0a}},\frac{h}{c_\text{stab}}), \quad  c_{stab}=c_{0w}\sqrt{\frac{\gamma_{a}\rho_{0w}}{\gamma_{w}\rho_\text{min}^{a}}},\\
\textcolor{black}{\Delta t_{\nu} = \text{min}(\frac{h}{\alpha c_{0w}}, \frac{h}{\alpha c_{0a}})}, \quad \Delta t_{ad} = \text{CFL}\frac{h}{\alpha_2c_{0w}},\\
  \Delta t\leq \min\left(\Delta t_{c}, \Delta t_{\nu}, \Delta t_{a}, \Delta t_{ad} \right),
  \end{cases}
\end{equation}
where $\rho_\text{min}^{a}$ is the minimum density of the gas phase and $\text{CFL}$ is the Courant-Friedrichs-Lewy number which is set to 1 in the present work. 
To ensure numerical stability, a '' stable sound sound speed'' $c_{stab}$ is considered for the time step following the work of \cite{sun2021accurate}. 

\textcolor{black}{
The time step $\Delta t_{\nu}$ associated with viscous effects is generally expressed as $\Delta t_{\nu} = 0.125 \text{min}(h^2 / \nu)$ (see e.g., \cite{colagrossi2016particle}). Since we are concerned about problems involving violent impacts, the Reynold number is generally high, and consequently, artificial viscosity is chosen to maintain the numerical stability.
By substituting the kinematic viscosity with $\nu = \alpha hc_0 /(2n +4)$ and the dimension $n =2$ in the present work, the $\Delta t_{\nu}$ is written as .
As indicated in Eq. \ref{eq:timestep}, $\alpha$ is relatively small, implying that the constraint on  $\Delta t_{\nu}$ is not the most restrictive.}

The $\Delta t_{ad}$ is related to the acoustic damper term, where the parameter $\alpha_2$ has to be large enough to achieve a smoothed pressure field similar to that solved with incompressible SPH. 
In singlephase problems, $\alpha_2$ typically ranges from 0 to 1 to ensure that $\Delta t_{ad}$ remains smaller than $\Delta t_{c}$ and $\Delta t_{a}$ \cite{sun2023inclusion}. 
\textcolor{black}{Notably, in present multiphase problems, the use of real gas sound speed $c_{0a} = 340$ m/s ensures gas physical compressibility, making $\Delta t_{c}$ the most restrictive factor. As a result, the parameter $\alpha_2$ can be as high as 10, further achieving a smoothed pressure field. In this way, a weakly compressible phase, like water, behaves more like an incompressible one due to the high acoustic damping, and the compressibility of the compressible phase, like gas, can also be well modeled with the physical sound speed.}

\subsection{Determination of the shifting velocity \texorpdfstring{$\delta\bm{u}$}- at different regions of the flow} \label{sec:shifting_velocity}
\textcolor{black}{In the $\delta^+$-SPH scheme \cite{sun2017deltaplus}, the shifting velocity is simply used to regularize the particle distribution, but not taken into account in the continuity and momentum equations. In contrast, in the present multiphase scheme, it is integrated into the governing equations rather than applied as an external displacement correction. To further clarify the effect of the two approaches, a comparison of the numerical results between these two approaches will be provided in Subsection \ref{section:dam-breaking}}.
The formula for shifting velocity is defined according to the work \cite{sun2019consistent} as follows:
\begin{equation}\label{eq:shifting-initial}
  \delta\bar{{\boldsymbol{u}}}_i=
-2h\text{Ma}c_0\sum\limits_{j}\nabla_i W_{ij}V_j,
\end{equation}
where Ma denotes the Mach number which is $U_{\text{max}} / c_0$.

However, the expression above has to be modified for particles in some special regions including free surface region $\Omega_{A}^{F}$ and solid walls $\Omega_{G}$ (see Figure \ref{fig:particle-set}). \textcolor{black}{For free surface region, only the tangential component of $\delta\bar{{\boldsymbol{u}}}_i$ is considered, similar to the work of Sun et al. \cite{SUN201725}. 
In this study, we introduce a novel treatment for $\delta \bm{u}$-terms at the fluid-solid interface, which will be discussed in the following subsection. This treatment allows us to set the shifting velocity $\delta \bm{u}$ of ghost particles to zero, differing from previous works that considered $\delta \bm{u}$ for ghost particles (see e.g., \cite{OGER201676}). This simplification avoids complex treatment on the ghost particles and further enhances computational efficiency. Furthermore, the stability of the implementation of the PST at the fluid-solid interface will be discussed and validation in Section \ref{section:results}.}
Therefore, the shifting velocity $\delta\bar{{\boldsymbol{u}}}_i$ is modified:
\begin{equation}\label{eq:shifting}
  \delta\bar{\boldsymbol{u}}_i=
  \begin{cases}
0 \,, &  \text{if } i \in\Omega_{G}, \\
\left(\bm{I}-\bm{n}_i\otimes \bm{n}_i\right)\delta\bar{\boldsymbol{u}}_i\,, & \text{if } i\in \Omega^{F}_{A},\\
\delta\bar{\boldsymbol{u}}_i\,, & \text{otherwise},
  \end{cases}
\end{equation}
where $\bm{I}$ denotes an identity matrix, and $\bm{n}_i$ represents the normal vector along the free surface of the $i$th particle. \textcolor{black}{Further details on the surface identification and normal vector computation can be found in \cite{marrone2010fast}.}
To further improve the numerical stability, a limit is imposed on the shifting term, so velocity deviation $\delta \boldsymbol{u}$ is finally expressed as:
\begin{equation}\label{eq:shifting-final}
  \delta{{\boldsymbol{u}}}_i=
  \begin{cases}
    \delta\bar{\boldsymbol{u}}_i\,, & \text{if } \|\delta\bar{\boldsymbol{u}}_i\| < 0.5{U_{\max}}\,\text{or} \,i\in \Omega^{F}_{A}\lor \Omega_{G} \\
0.5U_{\max}\frac{\delta\bar{{\boldsymbol{u}}}_i}  {\|\delta\bar{\boldsymbol{u}}_i\|}\,,  & \text{otherwise}
  \end{cases}
\end{equation}

\subsection{Extension of the consistent $\delta^{+}$-SPH model for multiphase flows: new treatments of \texorpdfstring{$\delta \bm{u}$}--terms} \label{section:interface-treatment}

In a multiphase system, particles at interfaces (i.e. fluid-solid and multi-fluid interfaces) often interact with neighboring particles that belong to different phases. Proper handling of these interactions is essential for maintaining stability at multiphase interfaces. Hence, for extending consistent $\delta^{+}$-SPH to multiphase flows, the treatment of $\delta\bm{u}$-terms is a key challenge.
To account for these interactions, we introduce the parameters $\varepsilon$, $\varphi$, and $\kappa$ to adjust the contribution of neighboring particles to the $\delta \boldsymbol{u}$-terms. We then rewrite Eq. \ref{eq:delta-uterms} as follows:
\begin{equation} \label{eq:delta-uterms-re}
  \fbox{\(\delta \bm{u}\)-terms}
    \begin{cases}
      \left\langle\nabla\cdot\left(\delta\boldsymbol{u}\right)\right\rangle_i=\sum\limits_{j}\left(\varepsilon_{ij}\delta\boldsymbol{u}_j-\varepsilon_{ii}\delta\boldsymbol{u}_i\right)\cdot\nabla_iW_{ij}V_j\\
      \left\langle\nabla\cdot\left(\rho\delta\boldsymbol{u}\right)\right\rangle_i=\sum\limits_{j}\left(\varphi_{ij}\rho_j\delta\boldsymbol{u}_j+\varphi_{ii}\rho_i\delta\boldsymbol{u}_i\right)\cdot\nabla_iW_{ij}V_j\\
      \left\langle\nabla\cdot\left(\rho\bm{u}\otimes\delta\bm{u}\right)\right\rangle_i=\sum\limits_{j}\kappa_{ij}\left(\rho_i\bm{u}_i\otimes\delta\bm{u}_i+\rho_j\bm{u}_j\otimes\delta\bm{u}_j\right)\cdot\nabla_iW_{ij}V_j.
    \end{cases}
  \end{equation}
  Here, the subscripts $ii$ and $ij$ of the parameters $\varepsilon$, $\varphi$, and $\kappa$ represent contributions from the particle $i$ itself and the neighboring particle $j$, respectively. The adoptions of $\varepsilon$ for the discretization of $\nabla\cdot\left(\delta\boldsymbol{u}\right)_i$, $\varphi$ for the discretization of $\nabla\cdot\left(\rho\delta\boldsymbol{u}\right)_i$ in the continuity equation; and $\kappa$ for the discretization of $\nabla\cdot\left(\rho\bm{u}\otimes\delta\bm{u}\right)_i$ in the momentum equation are introduced in the following subsections.

  \subsubsection{Numerical treatments of the \texorpdfstring{$\delta \bm{u}$}--term in the momentum equation}

  The $\delta \bm{u}$-term in the momentum equation is computed only when the particles $i$ and $j$ both belong to the same fluid phase. This means that the computing domain of the term $\nabla\cdot\left(\rho\bm{u}\otimes\delta\bm{u}\right)$ is restricted to a specific fluid phase. Consequently, the parameter $\kappa_{ij}$ is selected as follows:
  \begin{equation}
    \begin{cases}
      \left\langle\nabla\cdot\left(\rho\bm{u}\otimes\delta\bm{u}\right)\right\rangle_i=\sum\limits_{j}\kappa_{ij}\left(\rho_i\bm{u}_i\otimes\delta\bm{u}_i+\rho_j\bm{u}_j\otimes\delta\bm{u}_j\right)\cdot\nabla_iW_{ij}V_j\\
    \kappa_{ij} =
    \begin{cases} 
      1, & \text{if } \Omega(i) = \Omega(j), \\
      0, & \text{if } \Omega(i) \neq \Omega(j).
    \end{cases}
  \end{cases}
  \end{equation}

 Thanks to the symmetric structure of $\left\langle\nabla\cdot\left(\rho\bm{u}\otimes\delta\bm{u}\right)\right\rangle_i$ and the discretization that it only considers the same phase, this term remains unchanged when swapping the indexes $i$ and $j$. \textcolor{black}{Consequently, this term in the momentum equation does not affect the overall linear momentum conservation of the system.}            

\subsubsection{Numerical treatments of the \texorpdfstring{$\delta \bm{u}$}--terms in the continuity equation}

Effective management of the $\delta \bm{u}$-terms within the continuity equation is essential to ensure the stability and accuracy of the pressure field.
Furthermore, within the multiphase scheme, the discontinuity of density at multiphase interfaces increases the complexity of this issue. So, in the present subsection, we focus on the discretization of the $\delta \bm{u}$-terms (Eq. \ref{eq:delta-uterms-re}) in the continuity equation. 
The strategies for handing $\delta \bm{u}$-terms at fluid-solid and multi-fluid interfaces are discussed separately, as detailed below:

\textbullet{} \textbf{Strategy for \texorpdfstring{$\delta \bm{u}$}--terms applied at the fluid-solid interface}

As described in Eq. \ref{eq:shifting}, the shifting velocity $\delta \bm{u}$ for the ghost particles in the solid wall is zero. 
In terms of the first $\delta \bm{u}$-term $\nabla\cdot\left(\delta\boldsymbol{u}\right)$ in the continuity equation, the parameter $\varepsilon$ is set to $\varepsilon_{ii} = \varepsilon_{ij}=1$ straightforwardly. 
However, in terms of the second $\delta \bm{u}$-term $\nabla\cdot(\rho\delta\bm{u})$, the numerical instability in the pressure field occurs if the density of solid particles is directly involved since the density of the ghost particles is spurious. 

To avoid such instability, a strategy is employed which restricts the computation of $\nabla\cdot(\rho\delta\bm{u})$ only between particle pairs ($i\leftrightarrow j$) of the same fluid-phase. In other words, for a fluid particle, the solid wall particle has no influence on the calculation of the term $\nabla\cdot(\rho\delta\bm{u})$.
This is achieved by setting $\varphi_{ii}=\varphi_{ij} =0$ when particle $i$ and $j$ are from different particle sets, that is, when $\Omega\left(i\right)\neq \Omega\left(j\right)$ at the fluid-solid interface.   
Accordingly, the strategy for selecting $\varepsilon$ and $\varphi$ at the fluid-solid interface is summarized as follows:
\begin{equation}\label{eq:swith-for-fluid-solidinterface}
  \begin{cases}
    \left\langle\nabla\cdot\left(\delta\boldsymbol{u}\right)\right\rangle_i=\sum\limits_{j}\left(\varepsilon_{ij}\delta\boldsymbol{u}_j-\varepsilon_{ii}\delta\boldsymbol{u}_i\right)\cdot\nabla_iW_{ij}V_j\\
    \left\langle\nabla\cdot\left(\rho\delta\boldsymbol{u}\right)\right\rangle_i=\sum\limits_{j}\left(\varphi_{ij}\rho_j\delta\boldsymbol{u}_j+\varphi_{ii}\rho_i\delta\boldsymbol{u}_i\right)\cdot\nabla_iW_{ij}V_j\\
 \begin{cases}
  \begin{aligned}
  \varepsilon_{ii}=1,\quad & \varepsilon_{ij}=1                              \\
 \varphi_{ii}=\varphi_{ij}& = \begin{cases} 1, & \text{if } \Omega(i) = \Omega(j), \\ 0, & \text{if } \Omega(i) \neq \Omega(j).
\end{cases}            
\end{aligned}
\end{cases}
\end{cases}
\end{equation}

In this strategy, the density of solid wall particles is excluded from the approximation of $\nabla\cdot(\rho\delta\bm{u})$, ensuring that the influence of solid particles does not introduce spurious effects in the pressure field. 

\textbullet{} \textbf{Strategy for \texorpdfstring{$\delta \bm{u}$}--terms applied at the multi-fluid interface}  

We propose an alternative strategy for discretizing the $\delta \bm{u}$-terms in the continuity equation at the multi-fluid interface, and the two $\delta \bm{u}$-terms are treated in the same manner. 
In this strategy, the contribution of particle $i$ across its entire support domain is always considered, while the influence of neighboring particles from other phases is ignored.
To implement this strategy, the following modification is introduced:
 
For the contribution from particle $i$ itself, both $\varepsilon_{ii} = 1$ and $\varphi_{ii} = 1$ are adopted, regardless the neighboring particle $j$ belongs to which kind of particle sets. 
For the contribution of particle $j$, when particle $i$ interacts with neighboring particle $j$ from a different particle set, it is ignored by setting $\varepsilon_{ij}=0$ and $\varphi_{ij} =0$. Furthermore, this strategy also avoids directly using the density of other phases in computing $\nabla\cdot(\rho\delta\bm{u})$ in the continuity equation.
Finally, the strategy for selecting $\varepsilon$ and $\varphi$ at the multi-fluid interface is summarized as follows:
\begin{equation}\label{eq:multi-fluid-interface}
  \begin{cases}
    \left\langle\nabla\cdot\left(\delta\boldsymbol{u}\right)\right\rangle_i=\sum\limits_{j}\left(\varepsilon_{ij}\delta\boldsymbol{u}_j-\varepsilon_{ii}\delta\boldsymbol{u}_i\right)\cdot\nabla_iW_{ij}V_j\\
    \left\langle\nabla\cdot\left(\rho\delta\boldsymbol{u}\right)\right\rangle_i=\sum\limits_{j}\left(\varphi_{ij}\rho_j\delta\boldsymbol{u}_j+\varphi_{ii}\rho_i\delta\boldsymbol{u}_i\right)\cdot\nabla_iW_{ij}V_j\\
 \begin{cases}
  \begin{aligned}
    \varepsilon_{ii}=1, \quad & \varepsilon_{ij}= \begin{cases} 1, & \text{if } \Omega(i) = \Omega(j), \\ 0, & \text{if } \Omega(i) \neq \Omega(j).\end{cases}      \\                      
    \varphi_{ii}=1, \quad & \varphi_{ij} = \begin{cases} 1, & \text{if } \Omega(i) = \Omega(j), \\ 0, & \text{if } \Omega(i) \neq \Omega(j).\end{cases}          
  \end{aligned}
\end{cases} 
\end{cases} 
\end{equation}

\subsection{Final form of the discrete equations of multiphase consistent $\delta^{+}$ SPH} \label{section:interface-treatment-sum}

As discussed previously, the consistent $\delta^{+}$-SPH model extended for multiphase flows requires careful treatment of terms involving the shifting velocity $\delta \bm{u}$, particularly at fluid-solid and multi-fluid interfaces.  
In order to provide a clear illustration of the present multiphase $\delta^{+}$-SPH model, we suppose a fluid particle $i \in \Omega_A$; the neighboring particle $j$ of it can belong to the same fluid phase \(\Omega_A\), different fluid phases \(\Omega_B\), or the solid phase \(\Omega_G\), see Figure \ref{fig:particle-set}. 

Table \ref{tab:f} details the contributions of neighboring particles from different particle sets to the discretization of $\delta \bm{u}$-terms.
By identifying the set to which each neighboring particle belongs, the parameters $\varepsilon$, $\varphi$, and $\kappa$ are determined.
Based on Table \ref{tab:f}, the $\delta \bm{u}$-terms of particle $i$ can be clearly computed at multi-fluid and fluid-solid interfaces, even for special cases where the fluid particle $i$ is neighboring solid wall particles, particles from different fluid sets and particles from the same set within its support domain simultaneously.
\begin{table}[H]
  \centering
  \caption{Final form of the multiphase $\delta^{+}$-SPH model with new treatments of the $\delta \bm{u}$-terms. 
    Supposing a fluid particle $i$ belongs to particle set $\Omega_A$, the neighboring particle $j$ of fluid particle $i$ thus can belong to particle sets \(\Omega_A\), \(\Omega_B\), or \(\Omega_G\).}
  \label{tab:f}
  \resizebox{\textwidth}{!}{%
  \begin{tabular}{c|ccc}
    \toprule
    \multicolumn{1}{c|} {Continuity equation} & \multicolumn{3}{c} {
    $\dfrac{\dif \rho_i}{\dif t} = -\rho_i\langle\nabla \cdot \boldsymbol{u}\rangle_i
    -\rho_i \underbrace{\langle\nabla \cdot \delta \boldsymbol{u}\rangle_i}_{\text{$\delta \bm{u}$-term}}
    +\underbrace{\langle\nabla \cdot (\rho \delta \boldsymbol{u})\rangle_i}_{\text{$\delta \bm{u}$-term}} + \delta h c_0\mathcal{D}_i$} \\
    \multicolumn{1}{c|}{Momentum equation} & \multicolumn{3}{c} {
    ${\rho_i}\dfrac{\dif \boldsymbol{u}_i}{\dif t} = -{\langle\nabla p\rangle_i} + \langle\nabla \cdot \mathbb{V}\rangle_i + {\rho_i}\boldsymbol{g} + \underbrace{\langle\nabla \cdot (\rho \boldsymbol{u} \otimes \delta \boldsymbol{u})\rangle_i}_{\text{$\delta \bm{u}$-term}} + \boldsymbol{F}^{ad}_{i}$} \\
    \multicolumn{1}{c|}{Discretization of $\delta \bm{u}$-terms} & \multicolumn{3}{c} {
      $\begin{cases}
      \left\langle\nabla\cdot\left(\delta\boldsymbol{u}\right)\right\rangle_i=\sum\limits_{j}\left(\varepsilon_{ij}\delta\boldsymbol{u}_j-\varepsilon_{ii}\delta\boldsymbol{u}_i\right)\cdot\nabla_iW_{ij}V_j\\
      \left\langle\nabla\cdot\left(\rho\delta\boldsymbol{u}\right)\right\rangle_i=\sum\limits_{j}\left(\varphi_{ij}\rho_j\delta\boldsymbol{u}_j+\varphi_{ii}\rho_i\delta\boldsymbol{u}_i\right)\cdot\nabla_iW_{ij}V_j\\
      \left\langle\nabla\cdot\left(\rho\bm{u}\otimes\delta\bm{u}\right)\right\rangle_i=\sum\limits_{j}\kappa_{ij}\left(\rho_i\bm{u}_i\otimes\delta\bm{u}_i+\rho_j\bm{u}_j\otimes\delta\bm{u}_j\right)\cdot\nabla_iW_{ij}V_j
    \end{cases}$ 
    } \\
    \midrule
    \makecell{If $i\in \Omega_A$, \\ set of neighboring particle $j$} & If $j \in \Omega_A$ & If $j \in \Omega_B$ & If $j \in \Omega_G$ \\ 
    
    Parameters of $\delta \bm{u}$-terms  & 
    $\begin{cases}
              \begin{array}{rl}
                \varepsilon_{ii}&=\varepsilon_{ij}=1 \\
                \varphi_{ii}&=\varphi_{ij} = 1 \\ 
                \kappa_{ij} &=1 
              \end{array}
        \end{cases}$&
    $\begin{cases}
      \begin{array}{rl}
        \varepsilon_{ii} &= 1, \quad \varepsilon_{ij} = 0, \\                      
        \varphi_{ii} &= 1, \quad \varphi_{ij} = 0, \\
        &\kappa_{ij} = 0
      \end{array}
\end{cases}$&
    $\begin{cases}
            \begin{array}{rl}
              \varepsilon_{ii} &= 1, \quad \varepsilon_{ij} = 1, \\                      
              \varphi_{ii} &= 0, \quad \varphi_{ij} = 0, \\
              &\kappa_{ij} = 0
            \end{array}
      \end{cases}$ \\
      \midrule
  \end{tabular}
  }
\end{table}

It should be noted that in this work the particle mass is assumed to be constant. Consequently, mass conservation can be effectively maintained in these strategies. 
Due to the careful treatment of the $\delta\bm{u}$-terms at interfaces, the present $\delta^+$ SPH model exhibits stable behavior without the need for additional numerical techniques, such as background pressure, to handle interface instability. Among the different strategies we have tried \textcolor{black}{(see more details in the \ref{section:appendix}.}), the strategies for discretization of $\delta\bm{u}$-terms proposed in subsection \ref{section:interface-treatment} are the most stable for the treatment of multiphase interfaces. 
Various test cases are presented to validate the performance of the current multiphase SPH model in terms of stability, convergence and accuracy in the following section. Further theoretical studies on the strategies for treating the $\delta\bm{u}$-terms at multiphase interfaces will be explored in future work.

\section{SPH results and discussions}\label{section:results}
The consistent $\delta^{+}$-SPH model extended for multiphase flows in the present work is validated through six cases: two-phase hydrostatic test, multiphase dam breaking, slamming of LNG tank insulation panels, wedge water entry, two-layer liquid sloshing, and sloshing with entrapped air pockets. 
These comprehensive cases also include accurately probing and validating air pressure and evaluating the capability of handling free surface in the last two cases.

\begin{figure*}
  \centering
  \subfigure{
		\includegraphics[width=0.45\linewidth]{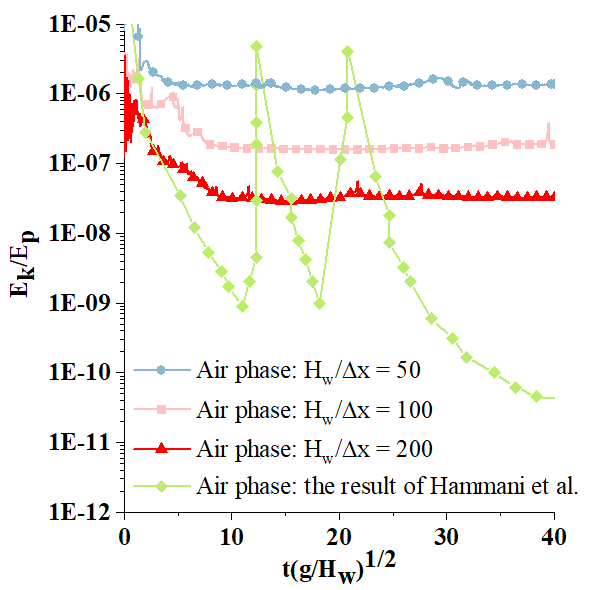}
  }
  \subfigure{
		\includegraphics[width=0.45\linewidth]{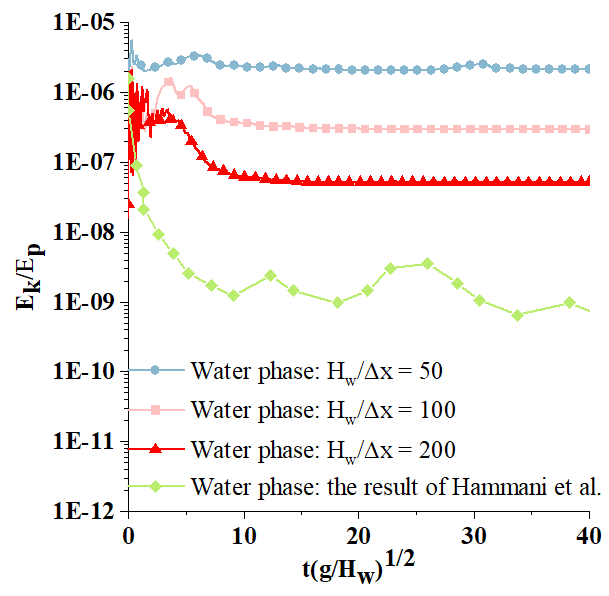}
  }
	\caption{ \textcolor{black}{The time history of the ratio between total kinetic energy $E_k$ and potential energy $E_p$ of the air phase (left) and the water phase (right) simulated with three particle resolutions, and compared with the result of Hammaini et al. \cite{hammani2020detailed}.}}
		\label{fig:engery-time-history-hydrostatic}
\end{figure*}

\begin{figure} 
  \centering
  \includegraphics[width=0.95\linewidth]{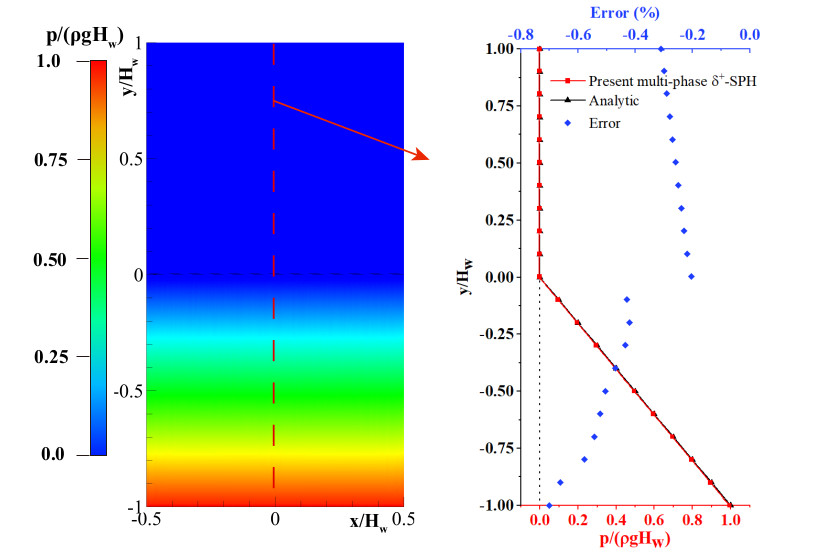}
  \caption{\textcolor{black}{The pressure field of two-phase hydrostatic at instant $t\left(g/H_w\right)^{1/2}=40$ (left), and the pressure distribution along the vertical centerline of the rectangular tank, compared with analytic trend (right). The numerical error, defined as $(p^{\text{SPH}}-p^{\text{Analytic}})/(\rho g H_w)$ is also plotted on the right side. The $\rho$ here is taken as water density $\rho = 1000 $ $\rm{kg/m^3}$.}} \label{fig:pressure-pressure-two-phase-hydrostatic}
\end{figure}

\subsection{Two-phase hydrostatic test}
Firstly, to assess the stability and robustness of the current multiphase SPH model, a two-phase hydrostatic scenario is simulated in a rectangular tank in this subsection.
The rectangular tank is of length $H_w$ and height $2H_w$ with the upper half filled with air and the lower half filled with water. The density of the water and air phase is set as 1000 $\rm{kg/m^3}$ and 1 $\rm{kg/m^3}$, respectively. 

\textcolor{black}{In this case, three particle resolutions $H_w / \Delta x = 50$, $H_w / \Delta x = 100$, and $H_w / \Delta x = 200$ are adopted, respectively. With the particle resolution increasing, the energy fluctuation at initial time due to the initial Cartesian distribution can effectively decrease. This issue has been discussed in \cite{COLAGROSSI20121641}, where a particle packing algorithm was proposed to reduce spurious initialization effects and improve numerical stability.}
As shown in Figure \ref{fig:engery-time-history-hydrostatic}, the disturbing kinetic energy of the two phases  decreases rapidly and converges close to zero with particle resolution $H_w / \Delta x = 200$, indicating that the present multiphase $\delta^{+}$-SPH model achieves good convergence with $H_w / \Delta x = 200$ and this particle resolution is chosen to discuss following. \textcolor{black}{Additionally, compared to the results from \cite{hammani2020detailed} simulated with the multiphase $\delta$-SPH model, the disturbing kinetic energy of the two phases in the present scheme converges more quickly due to the effect of PST. However, the use of PST can contribute some kinetic energy, as long as the particle distribution is not uniform during the simulation. As a result, the settled energy is slightly higher than the result simulated by Hammani et al. \cite{hammani2020detailed}. Furthermore, the adoption of background pressure in the work of Hammani et al. \cite{hammani2020detailed} can also lead to some numerical dissipation \cite{sun2018multi}, contributing to a lower kinetic energy.
However, the energy ratio of $10^{-8}$ with particle resolution $H_w / \Delta x = 200$ is sufficiently low to confirm the stability of the present multiphase scheme.} 

\textcolor{black}{At $t\left(g/H_w\right)^{1/2}=40$, as illustrated on the left side of Figure \ref{fig:pressure-pressure-two-phase-hydrostatic}, the pressure distribution of water and gas phases remains stable. On the right side of Figure \ref{fig:pressure-pressure-two-phase-hydrostatic}, the pressure distribution along the centerline of the tank is compared with the analytical trend plotted along the bottom axis, and the numerical error calculated by $(p^{\text{SPH}}-p^{\text{Analytic}})/(\rho g H_w)$ is plotted along the top axis. As one can observe, the pressure gradient near the water-air interface is accurately captured, with the numerical error as low as $-0.8\%$, demonstrating that the present model maintains stable behavior at the multi-fluid interface over long-term simulations as well.} 

\begin{figure} 
  \centering
  \includegraphics[width=0.85\linewidth]{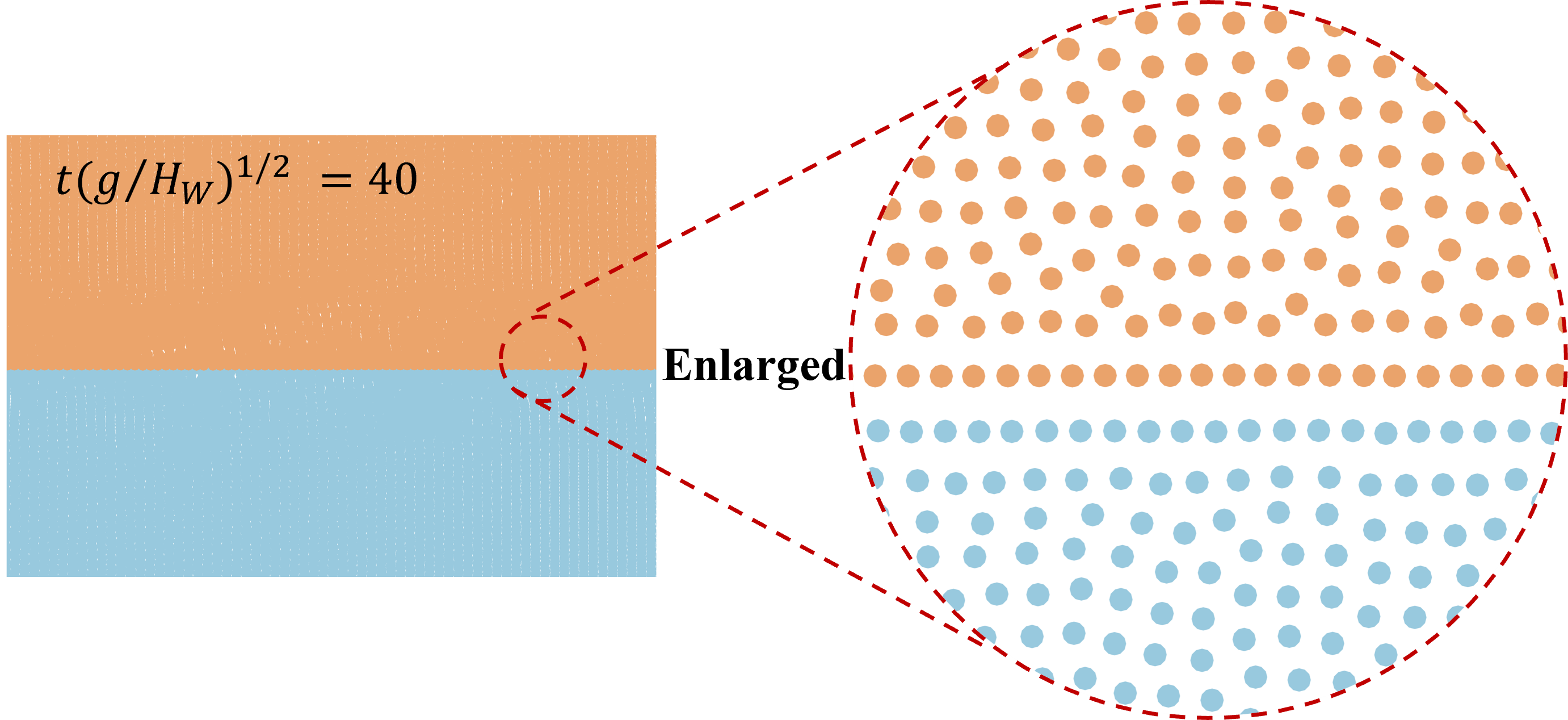}
  \caption{The particle distribution of two-phase hydrostatic at instant $t\left(g/H_w\right)^{1/2}=40$.} \label{fig:particle-distribution-two-phase-hydrostatic}
\end{figure}

Furthermore, as shown in Figure \ref{fig:particle-distribution-two-phase-hydrostatic}, the particles are uniformly distributed and no interpenetration occurs at the water-air interface during the extended simulation period without the need of background pressure.

\subsection{Dam-break flows impacting a solid wall}\label{section:dam-breaking}

\begin{figure}
  \centering
  \includegraphics[width=0.85\linewidth]{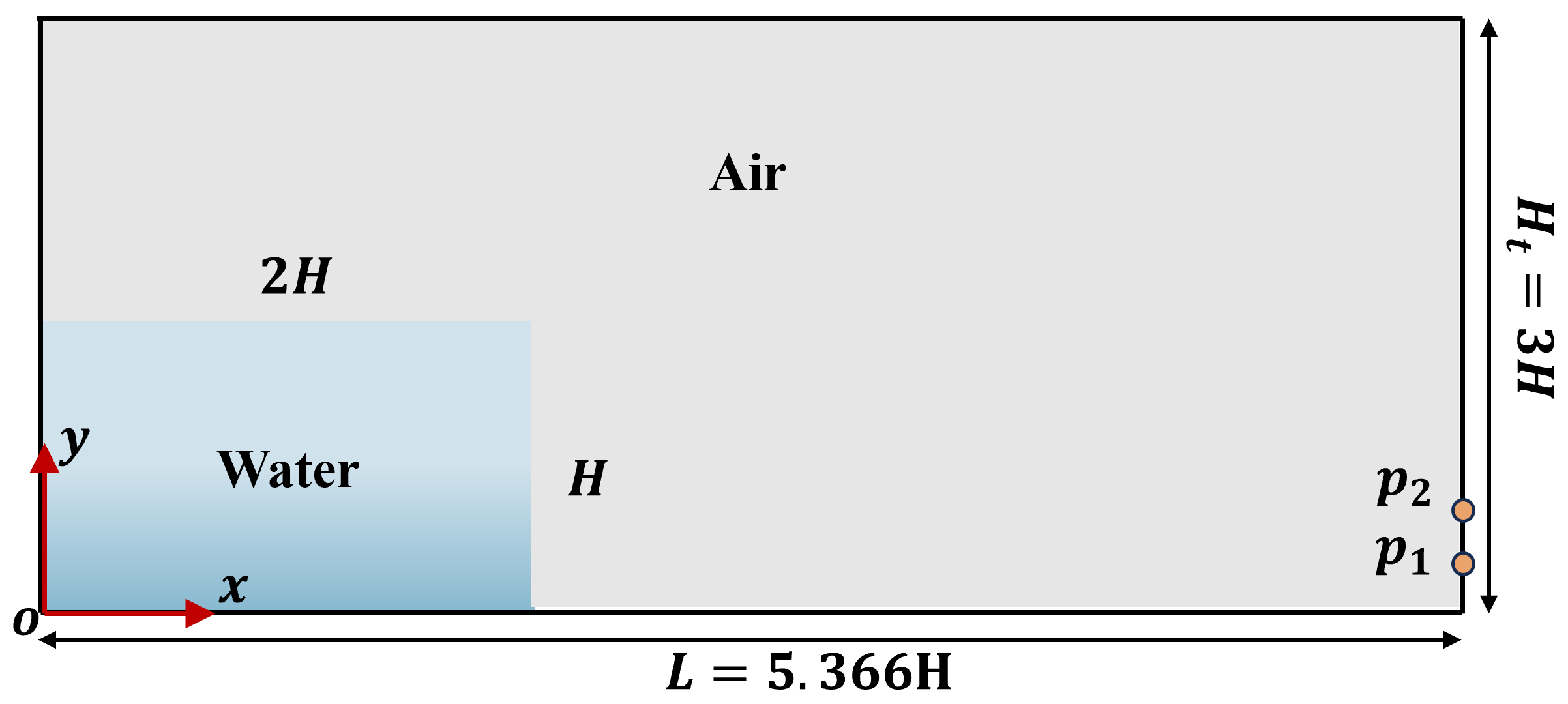}
  \label{fig:11}
  \caption{
    The dam breaking scenario involves a water body with a height of $H$ and a width of $2H$. The remaining region is filled with the air phase.} \label{fig:dam-breaking-diagram}
\end{figure}

It is well-known that dam breaking involves surface breaking and violent impact phenomena, being a classic benchmark (see \cite{he2022stable,colagrossi2003numerical} as examples) in the SPH community to validate the accuracy and stability of SPH models.
In this Subsection, water-air phase dam breaking impacting a solid wall simulated with the $\delta^{+}$-SPH model proposed by Sun et al. \cite{sun2017deltaplus} and the present multiphase $\delta^{+}$-SPH model is discussed. 
Figure \ref{fig:dam-breaking-diagram} shows the geometrical configuration and the initial pressure distribution of the water phase can be seen in \cite{zhou1999nonlinear}. The height of the water body is $H = 0.6$ $\rm{m}$ and the width $2H = 1.2$ $\rm{m}$. 
Additionally, the experimental sensors, each with a diameter of $d = 90$ mm, are located $y_1 = 160$ mm and $y_2 = 584$ mm away from the bottom wall, respectively. Accordingly, the numerical probes $p_1$ and $p_2$, are placed at the bottom of the experimental sensors at $y_1 = 115$ mm and $y_2 = 539$ mm, respectively.

\begin{figure}
  \centering
  \includegraphics[width=0.95\linewidth]{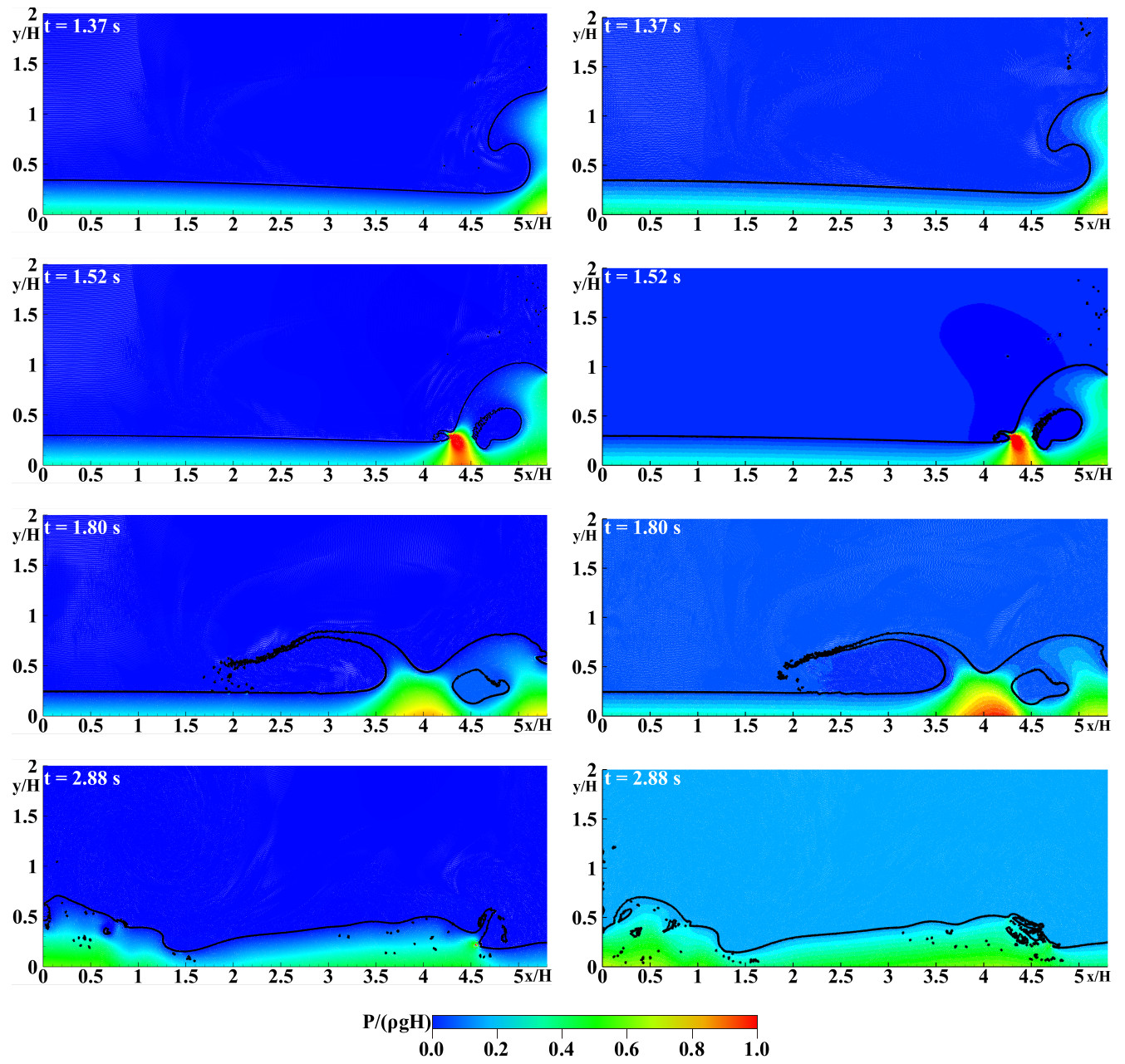}
  \caption{The results of pressure field simulated with the present multiphase $\delta^{+}$-SPH model(left column) and $\delta^{+}$-SPH proposed by Sun et al. \cite{sun2017deltaplus} (right column) for multiphase flows.} \label{fig:dam-breaking-comparison-between-delta-plus-and present}
\end{figure}

\begin{figure}
  \centering
  \includegraphics[width=0.95\linewidth]{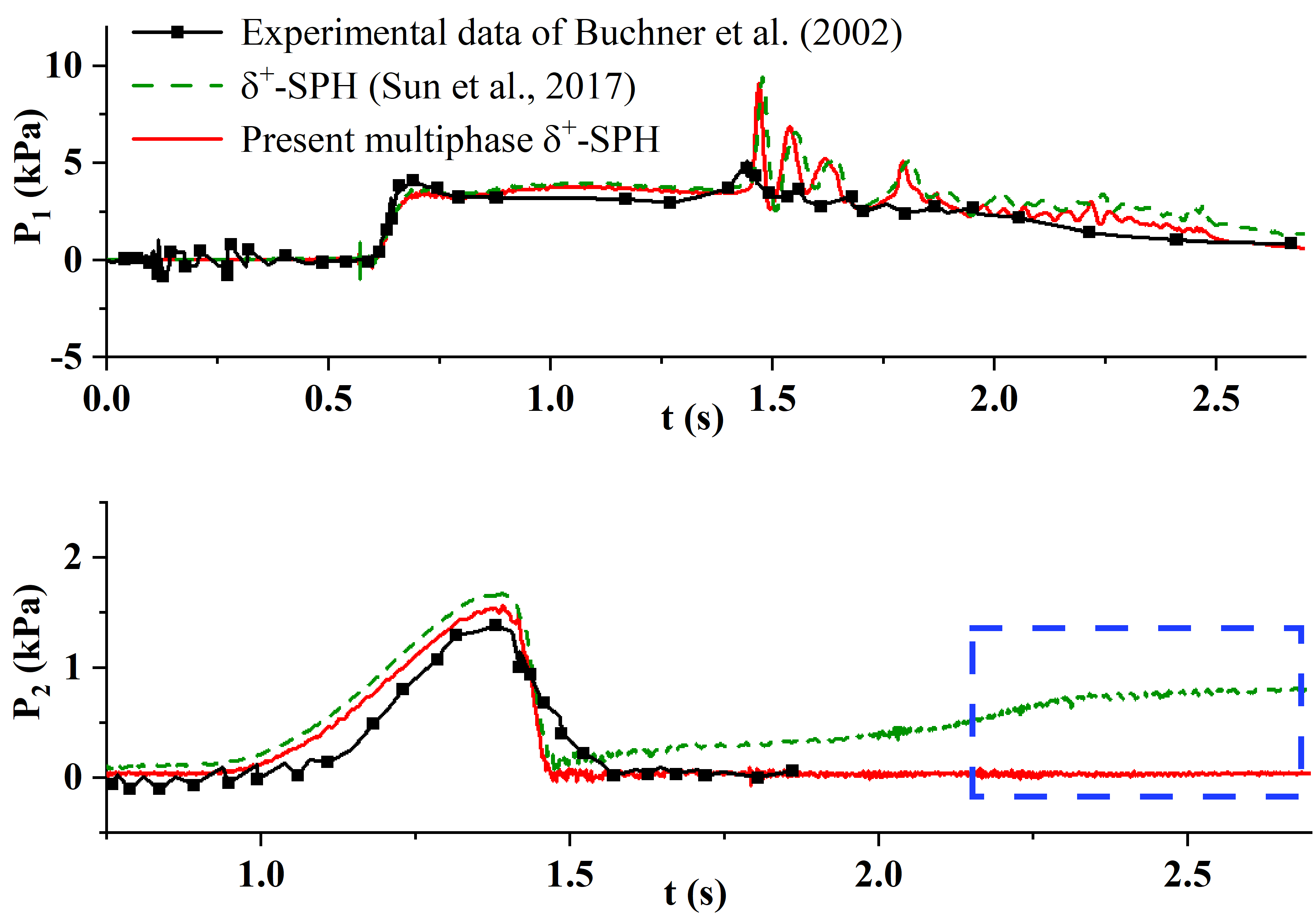}
  \caption{The evolution of pressure monitored at $p_1$ (top) and $p_2$ (bottom), simulating by $\delta^{+}$-SPH model with simple particle shifting \cite{sun2017deltaplus} and the present multiphase $\delta^{+}$-SPH model compared with experimental data \cite{buchner2002green}.} \label{fig:dambreaking-pressure-verification}
\end{figure}

Figure \ref{fig:dam-breaking-comparison-between-delta-plus-and present} depicts the pressure fields and water shapes simulated by the $\delta^{+}$-SPH model with simple particle shifting \cite{sun2017deltaplus} and the present multiphase $\delta^{+}$-SPH model at several instants. 
Observing the shape of water at $t = 1.37 $s, the water impacts the downstream wall and climbs along the solid wall. Furthermore, more water particles splash into the air region, as observed at $t = 1.52$ s. These evolutions of impact flows can be accurately captured by the two multiphase SPH models.
According to the experimental setup, water and outside air connected to the atmosphere are essentially incompressible, whereas the entrapped air pocket undergoes compression and expansion.
However, the results show a noticeable discrepancy that the air pressure simulated by the ${\delta^{+}}$-SPH model \cite{sun2017deltaplus} increases significantly during the impact state.
\textcolor{black}{This is because in the $\delta^{+}$-SPH model proposed by Sun et al. \cite{sun2017deltaplus}, the shifting velocity is simply used to regularize the particle distribution, but not considered in the discrete form of the governing equations, which may lead to the non-conservation of volume and therefore leads to the increase of the background pressure (green curve)  in Figure \ref{fig:dambreaking-pressure-verification}.} To alleviate the above issue, the present work focuses on treating the $\delta \bm{u}$-terms at the multiphase interfaces, the results of which, are well illustrated by the pressure contour shown on the left side of Figure \ref{fig:dam-breaking-comparison-between-delta-plus-and present} and the evolution of  pressure (red curve) predicted in Figure \ref{fig:dambreaking-pressure-verification}.

Comparisons of the pressure of the two sensors between the experimental \cite{buchner2002green} and numerical results are shown in Figure \ref{fig:dambreaking-pressure-verification}.
The results of the SPH simulation are in good agreement with the experimental data \cite{buchner2002green} before water overturning, especially at the sensor $p_2$. As shown in Figure \ref{fig:dam-breaking-comparison-between-delta-plus-and present} at $t = 1.52$ s, air entrainment occurs near the right wall, leading to compression and expansion of the entrapped air. Consequently, the water near the pressure sensor $p_1$ experiences excitation, and results in pressure oscillations that can be observed in Figure \ref{fig:dambreaking-pressure-verification}. However, these behaviors are not visible with the experimental sensors due to 3D effects. 

\begin{figure}
  \centering
  \includegraphics[width=0.95\linewidth]{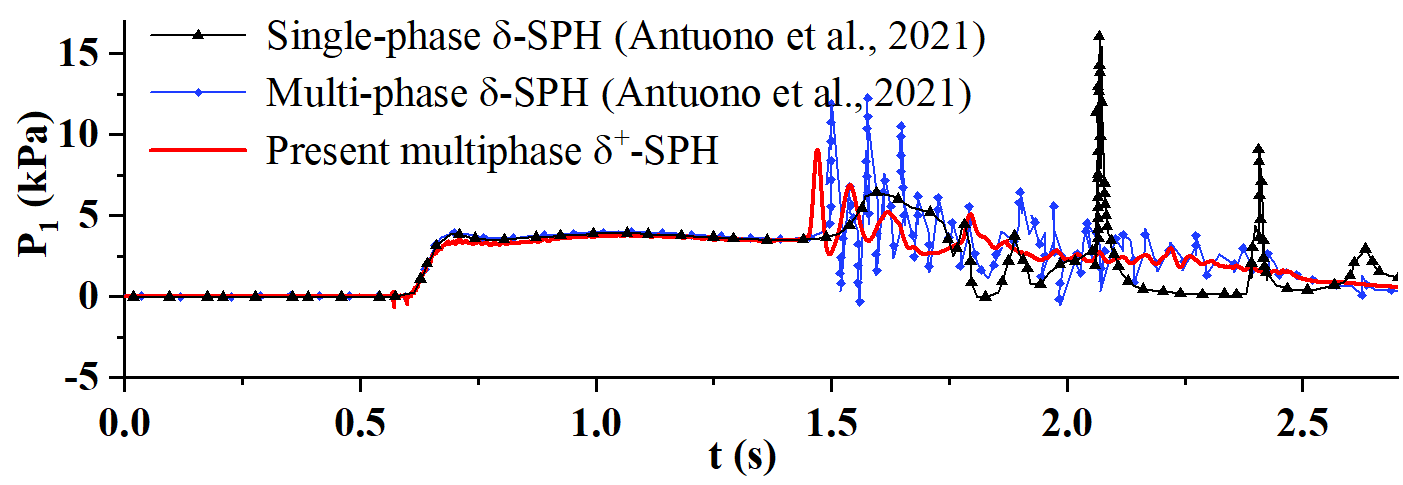}
  \caption{Comparison of the evolution of pressure monitored at $p_1$ between single- and multiphase $\delta$-SPH given by Antuono et al. \cite{antuono2021violent} and the present multiphase $\delta^{+}$-SPH model.} \label{fig:dambreaking-p1-compared-Antuono}
\end{figure}

Furthermore, we compare the time history of pressure at the point $p_1$ between single- and multiphase $\delta$-SPH results of Antuono et al. \cite{antuono2021violent} in Figure \ref{fig:dambreaking-p1-compared-Antuono} with the 2D SPH simulations.
It is observed that the aforementioned pressure oscillations are captured in multiphase simulations, while the single-phase simulation fails due to the lack of consideration of the air phase. Notably, the pressure oscillations are more periodic and less noisy when simulated with the present multiphase $\delta^{+}$-SPH.
The accuracy of the air phase pressure obtained by the present multiphase $\delta^{+}$ SPH will be qualitatively validated in the following subsections.

\begin{figure}
  \centering
  \includegraphics[width=0.95\linewidth]{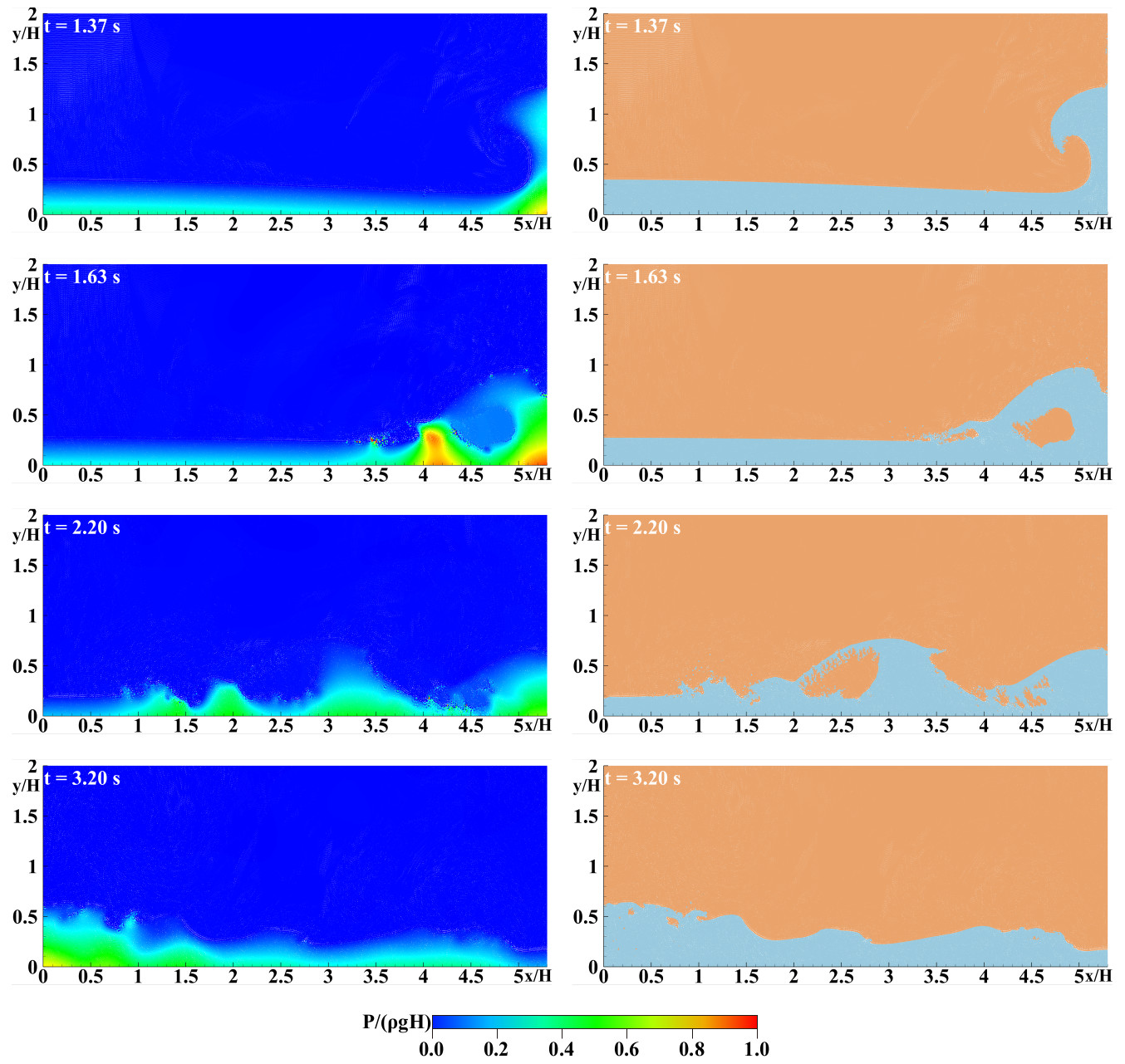}
  \caption{The pressure field (left) and particle distribution (right) simulated with artificial viscosity parameter $\alpha = 0.02$.}\label{fig:dambreaking-artificial-}
\end{figure}

\begin{figure}
  \centering
  \includegraphics[width=0.95\linewidth]{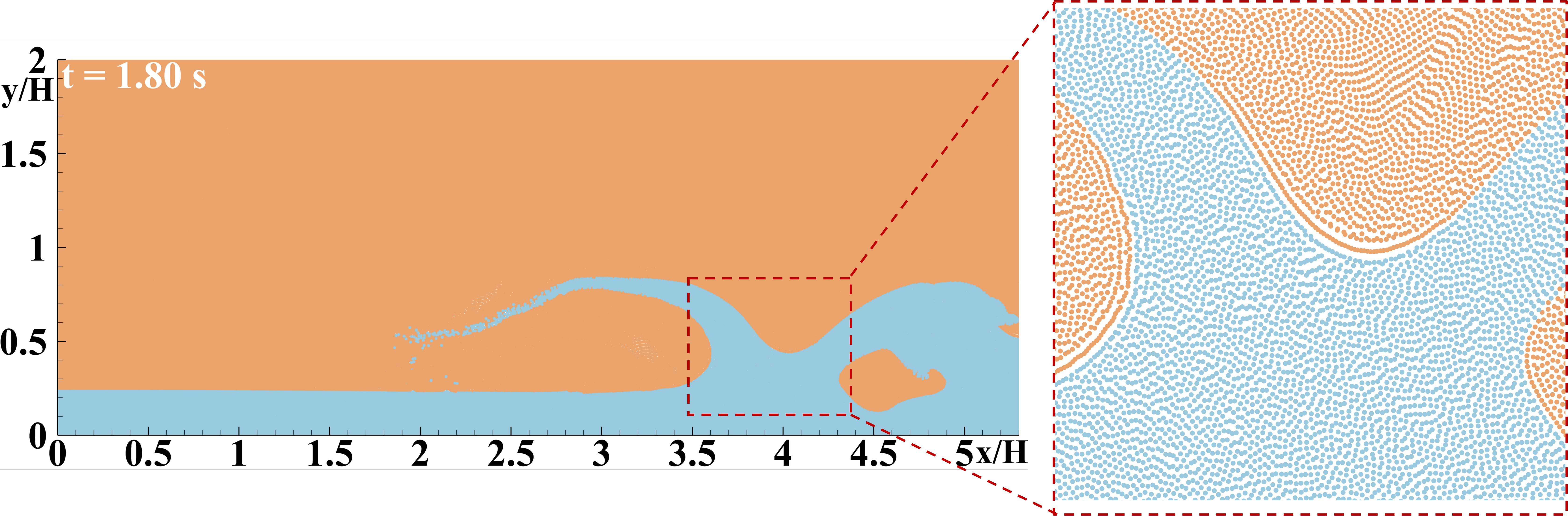}
  \caption{Particle distribution of dam breaking problem simulated by present consistent multiphase $\delta^+$-SPH model with $\alpha = 0.1$. Zoom-in view of the flow region near the interface with large deformation.} \label{fig:dambreaking-particle-ditribution}
\end{figure}

\textcolor{black}{As mentioned earlier, in the present multiphase scheme, the interface stability achieved is free from background pressure and repulsive force methods. Instead, PST helps to maintain the stability of interfaces. To better understand the effects of artificial viscosity and PST, we also simulate the dam-breaking case with artificial viscosity parameter $\alpha = 0.02$. As seen in Figure \ref{fig:dambreaking-artificial-}, the pressure field is also smoothed and particle distribution is uniform, despite some minor splashing observed at $t =1.63$ s. Particularly, at longer times, such as $t = 3.2$ s, the physical separation between the gas and water particles is naturally achieved, with the gas particles as light phase staying on top and the water particles as dense phase remaining below. This suggests that artificial viscosity is not the primary factor in ensuring interface stability. For the case with $\alpha = 0.1$,
Figure \ref{fig:dambreaking-particle-ditribution} shows that the particle distribution does not exhibit penetration into the phase interface. Importantly, even in cases of air entrapped and large interface deformation, the present SPH model simulated with either $\alpha = 0.1$ or $\alpha = 0.02$, effectively preserves a smooth interface, free from non-physical gaps. However, less splashing is observed with $\alpha = 0.1$. Therefore, to obtain a clearer interface for violent impact problems, $\alpha = 0.1$ is selected in the present work. Furthermore, the numerical results without particle splashing are similar between the two values of $\alpha$, which will be further discussed in Subsection \ref{section:two-layer}.}

\textcolor{black}{Moreover, in the present work, we highlight that the proper implementation of the PST is the key factor in stabilizing the interfaces rather than relying on the use of $\alpha = 0.1$. As demonstrated in the \ref{section:appendix}, improper handling of the PST at multiphase interfaces can lead to noise pressure and inter-penetration, potentially causing simulation failure despite using $\alpha = 0.1$. Thanks to the proper strategies for the implementation of the PST in Section \ref{section:interface-treatment}, interpenetration is successfully avoided.
}

\subsection{Slamming of LNG tank insulation panel}
In this subsection, in order to validate the capability of the present multiphase SPH model to handle fluid-solid interaction problems, and following the experiment of Marrone et al. \cite{marrone2017challenges}, a complex problem of LNG tank insulation panel slamming is studied, where the effect of air cushion should not be neglected. So in this case, the ability of the present multiphase scheme to model air compressibility can be further validated. It is notable that in this case, the air is compressed into a very small volume and with a high-peak pressure, which may cause nonphysical interface penetration.
A sharpness force is applied in the momentum equation in Eq. \ref{eq:ale-governing}. This force helps preserve interface clarity in this particular case, while in other scenarios studied in this work, interface stability is maintained without the need for the sharpness force. The interface sharpness force 
$F_i^{sharpness}$ \cite{grenier2009hamiltonian} added to the momentum equation is written as
\begin{equation}
F_i^{sharpness}=-\chi\sum_{j\,\notin \,\Omega_G,\,\atop \Omega\left(j\right)\neq \Omega\left(i\right)} ( \|p_i\|\,+\, \|p_j\|\,)\cdot \nabla_i W_{ij} \,V_j\,,
\end{equation}
where $\chi =0.08$ is adopted.

\begin{figure}
  \centering
  \includegraphics[width=0.9\linewidth]{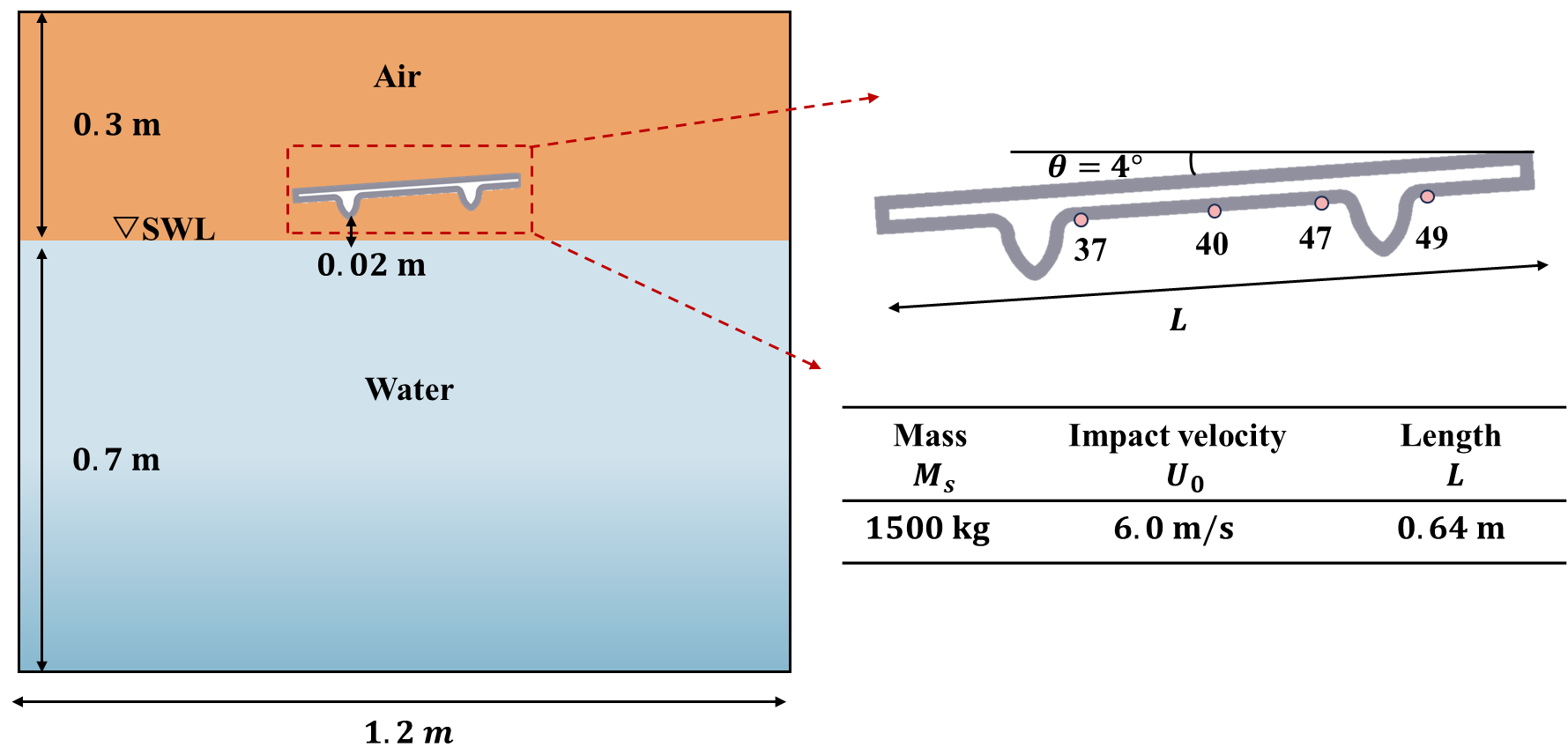}
  \caption{The numerical setup and diagram of 2D profile of the LNG tank insulation panel. Its length is $L = 0.64$ m and its mass is  $M_s = 1500$ kg. Four pressure monitor points 37, 40, 47, 49 are placed on the panel.} \label{fig:LNG_water_entry_diagram}
\end{figure} 

The insulation panel with a mass of $M_s = 1500$ kg falls with an impact velocity of $U_0 = 6.0$ m/s. The particle resolution is $L/\Delta x = 320$. The initial configuration and geometric characteristics of the panel are illustrated in Figure \ref{fig:LNG_water_entry_diagram}. The initial configuration of the LNG profile is set at a height of 0.02 m above the free surface, in order to account for the air effect. The sound speed of the water is set at a value of $c_{0w} = 212$ m/s. 

After $0.0036$ s, the left dent of LNG tank insulation panel reaches water free surface. This moment is denoted as $ t = -0.0036$ s, so the time when the panel touches the free surface is defined as $t = 0$ s.  
 \begin{figure*}
  \centering
  \subfigure{
    \includegraphics[width=0.45\linewidth]{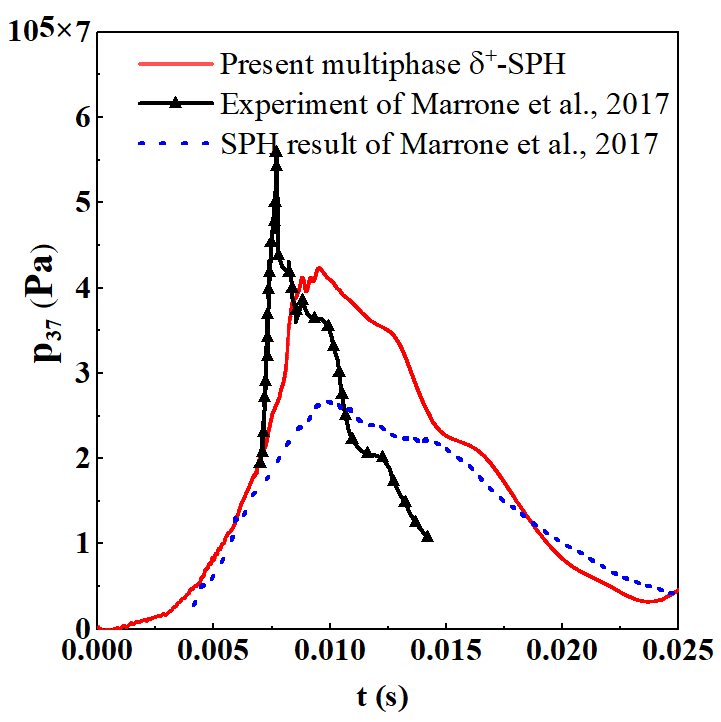}\label{fig:p37}
    \includegraphics[width=0.45\linewidth]{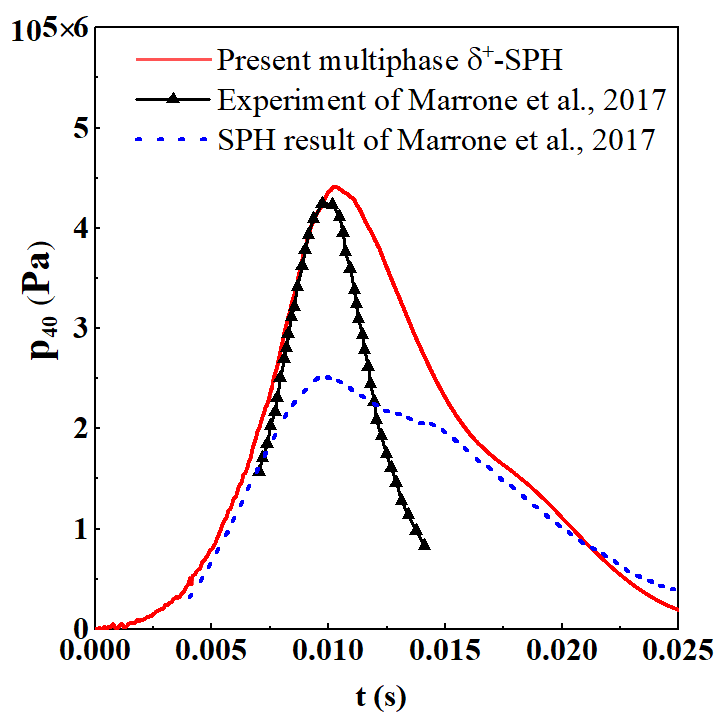}\label{fig:p40}
  }
  \subfigure{
    \includegraphics[width=0.45\linewidth]{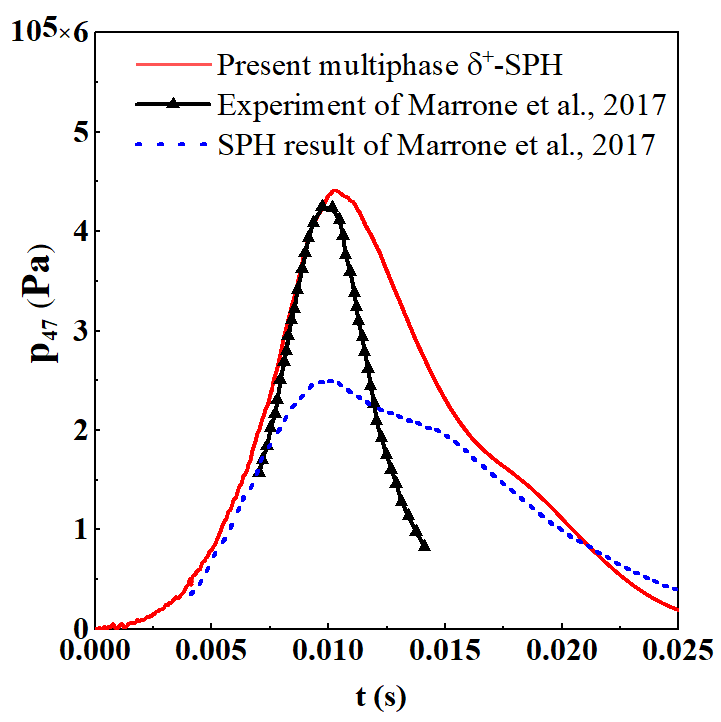}\label{fig:p47}
    \includegraphics[width=0.45\linewidth]{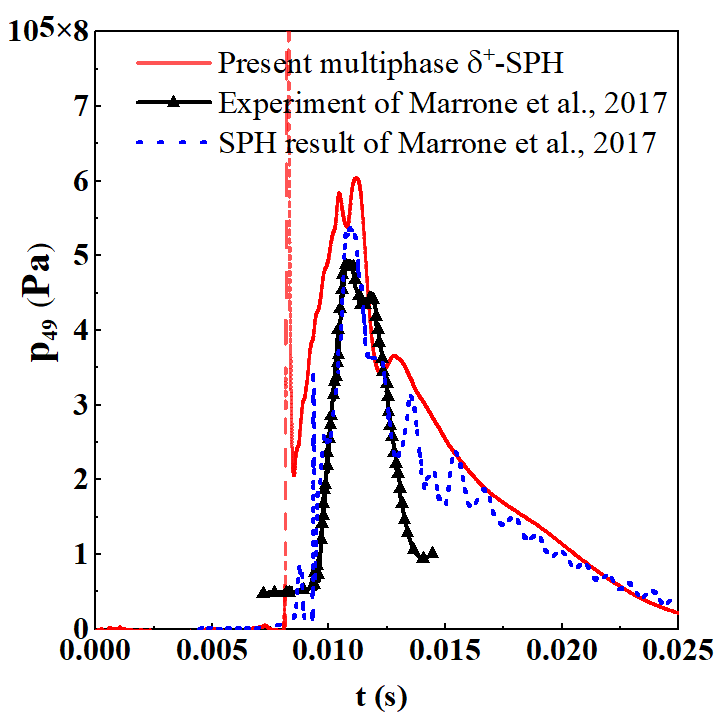}\label{fig:p49}
  }
  \caption{The evolution of pressure compared with experimental data and SPH results of Marrone et al. \cite{marrone2017challenges} probed at monitor points 37, 40, 47, 49.}
  \label{fig:LNG-panel-pressure}
\end{figure*}
Figure \ref{fig:LNG-panel-pressure} illustrates the evolution of pressure at four monitoring points designated as 37, 40, 47, and 49, respectively. 
The pressure loads of the four points predicted by the present multiphase SPH generally agree with the experimental data given in \cite{marrone2017challenges}. 
Compared to the numerical results obtained by Marrone et al. \cite{marrone2017challenges} using the Riemann-ALE SPH scheme, the pressure peaks in this study are larger and closer to the experimental data, particularly at points 40 and 47, where the pressure peaks are strongly associated with air bubble oscillations.  
This can be attributed to the accurate modeling of air compressibility using physical sound speed in the present multiphase scheme. 

\begin{figure}
 \centering
 \includegraphics[width=0.9\linewidth]{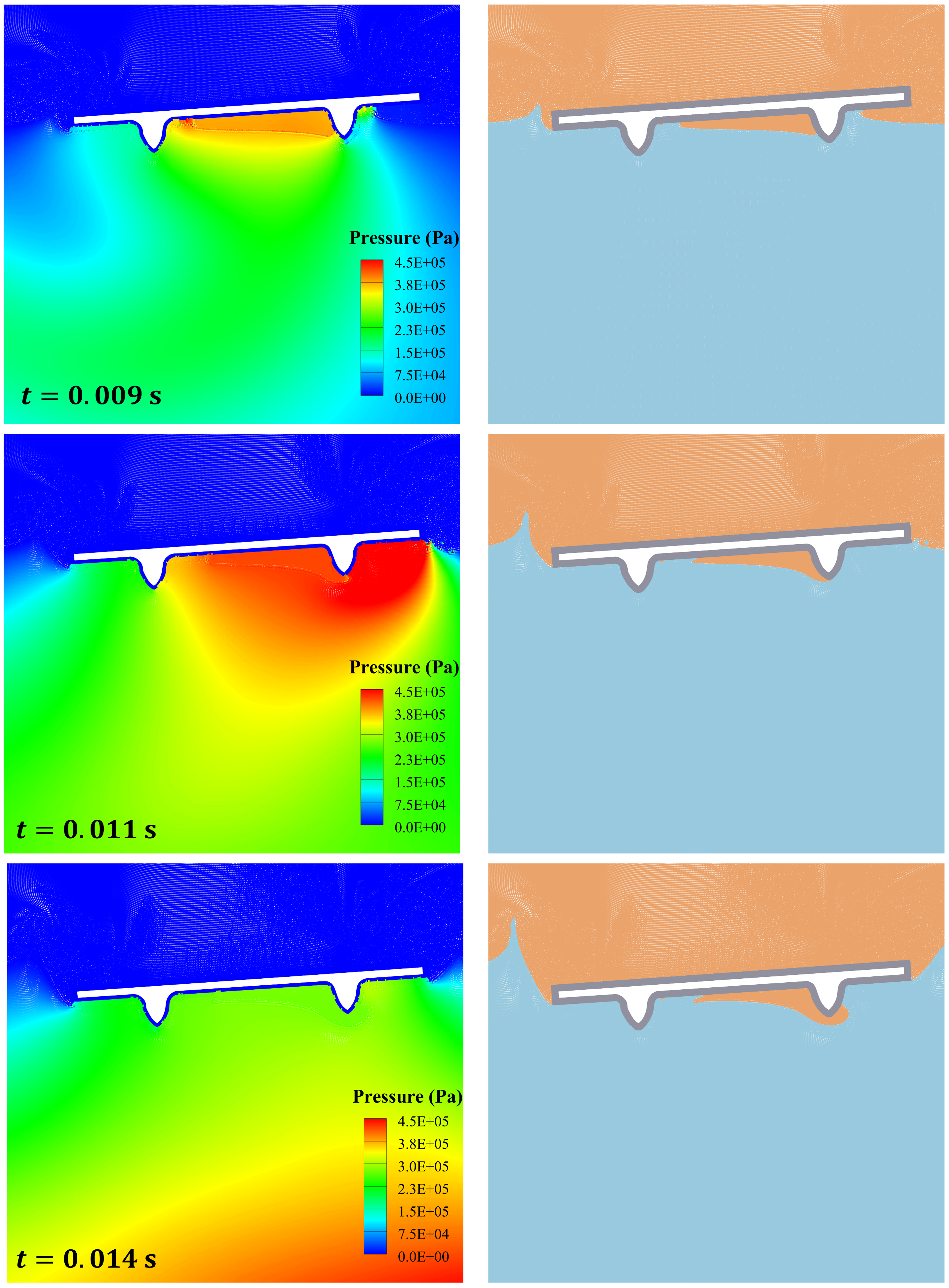}
 \caption{The pressure field (left) and particle distribution (right) at instants when air is entrapped.} \label{fig:panel-pressure-itype} 
\end{figure}

Typical instants of air entrapped are given in Figure \ref{fig:panel-pressure-itype}. 
As the panel falls, air is entrapped and compressed seeing at instants $t = 0.009$ s and $t = 0.011$ s. 
Subsequently, the pressure of air rises and peaks between $t = 0.009$ s and $t = 0.014$ s, which can also be observed in the pressure history at points 40 and 47 in Figure \ref{fig:LNG-panel-pressure}.    
As air escapes from the right dent at $t = 0.014$ s, the air bubble expands and the pressure decreases. 

 \subsection{The whole process of wedge water entry}
\begin{figure}
   \centering
   \includegraphics[width=0.9\linewidth]{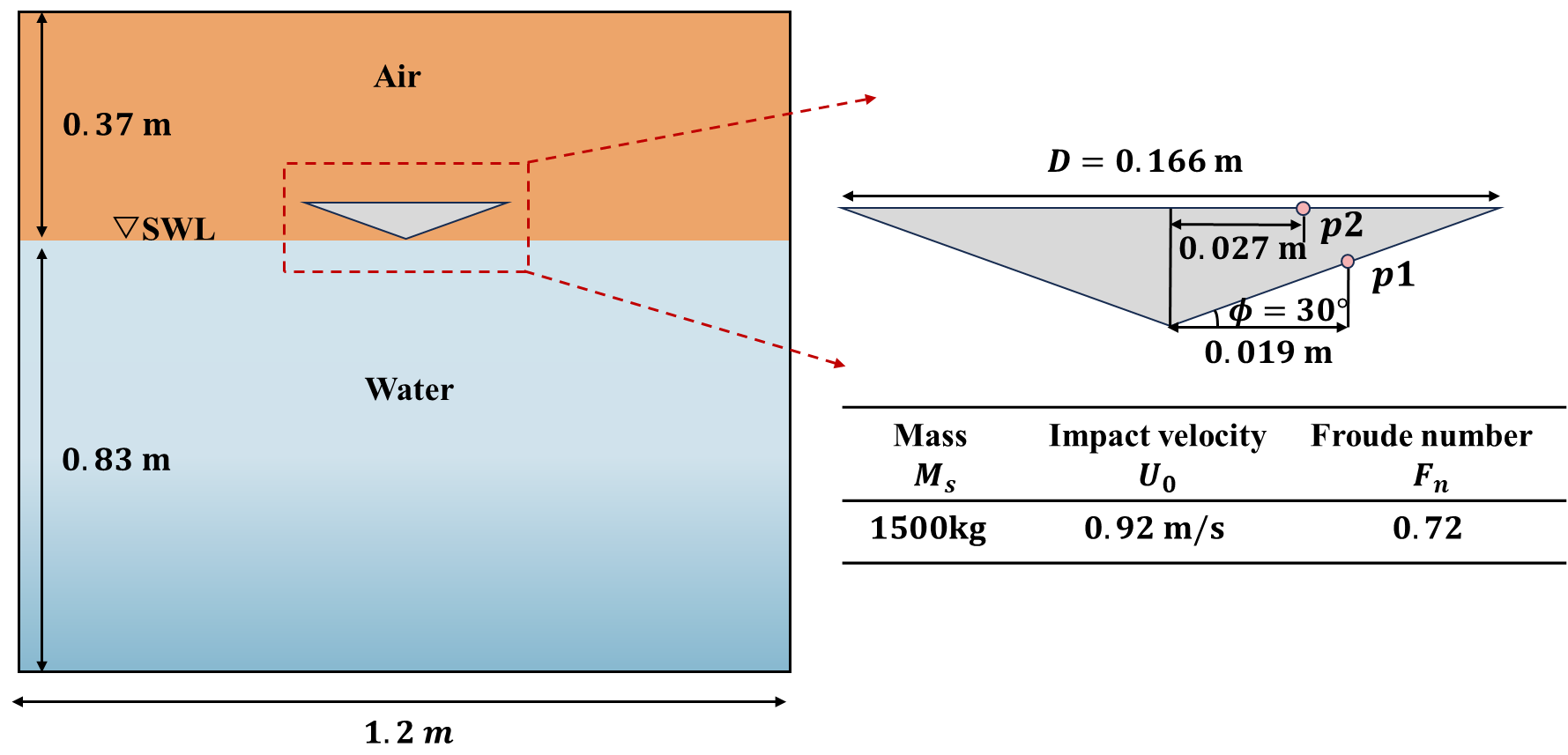}
   \caption{The diagram of the initial numerical setup. The wedge is simplified into an isosceles triangle structure with \textcolor{black}{a dead-rise angle $\phi = 30^\circ $}. Its length is $D = 0.166$ m and its mass is  $M_s = 32.3$ kg. Two pressure monitors placed at $p1$ and $p2$ on the wedge are used to probe the pressure at the water surface and the air surface, respectively.} \label{fig:water_entry_diagram}
 \end{figure}

 According to the experiment of Wang et al. \cite{wang2015experimental}, the whole process of water entry is simulated with particular attention to the state of cavity closure in this subsection. A wedge with mass $M_s  = 32.3 $ kg penetrating water surface with an impact velocity $U_0=0.92$ m/s and the Froude number $F_n = U_0\sqrt{gD}$ = 0.72 is studied. The initial setup and shape parameters of the wedge are shown in Figure \ref{fig:water_entry_diagram}. 
 
\begin{figure}[htbp]
    \centering
    \begin{minipage}{0.49\linewidth}
      \centering
      \includegraphics[width=0.9\linewidth]{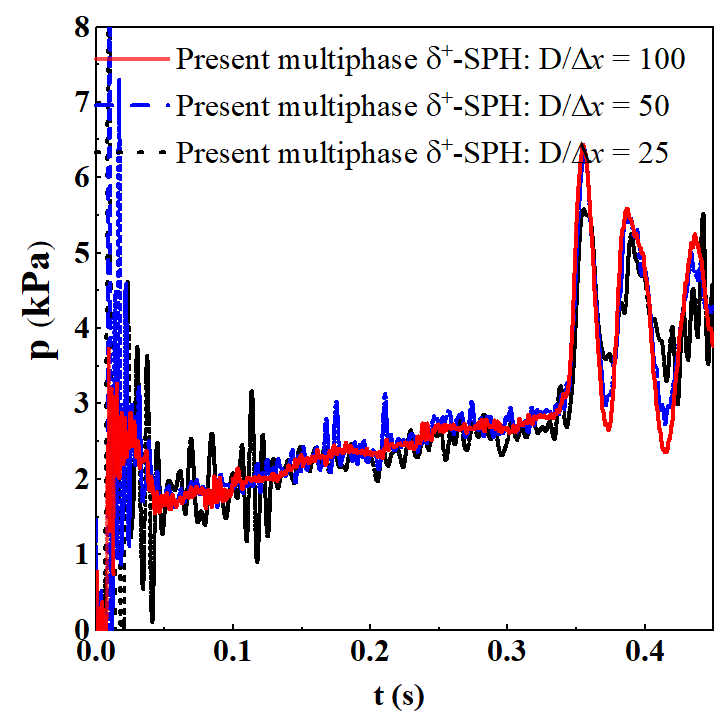}
      \caption{The evolution of pressure probed at the $p1$ simulated with particle resolutions $D/\Delta x = 25$, $D/\Delta x = 50$ and $D/\Delta x = 100$.}
      \label{fig:water-entry-convergence}
    \end{minipage}
    \begin{minipage}{0.49\linewidth}
      \centering
      \includegraphics[width=0.9\linewidth]{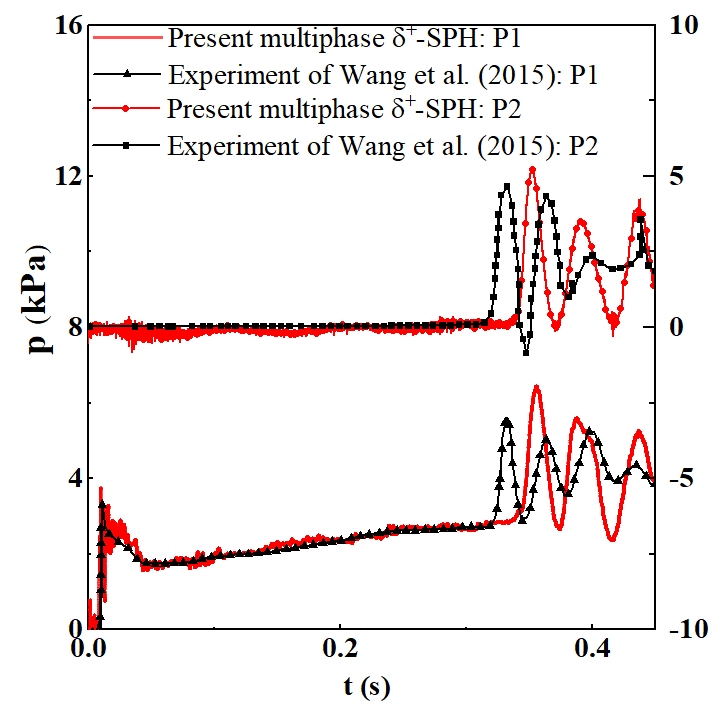}
      \caption{The verification of the pressure evolution between the present SPH results and experiment data. \cite{wang2015experimental} }
      \label{fig:water-entry-verification}
    \end{minipage}
  \end{figure}

 Firstly, the convergence of particle spacing is conducted. Three particle resolutions related to the width of the wedge $D$ respectively are $D/\Delta x = 25$, $D/\Delta x = 50$ and $D/\Delta x = 100$.
 As plotted in Figure \ref{fig:water-entry-convergence}, the evolution of the pressure probed at $p1$ converges when the particle resolution reaches $D/\Delta x = 50$ and $D/\Delta x = 100$. Therefore, the particle resolution $D/\Delta x = 100$ is employed in the following discussion in this subsection. 
 
 The simulation results of the whole process of wedge water entry including the pressure probed on the impact side $p1$ and the upper wedge surface $p2$ are verified with the experimental results shown in Figure \ref{fig:water-entry-verification}. 
 The pressure value probed at $p1$ is shown on the left axis and $p2$ on the right axis of the graph.
 As can be seen, the pressure at $p1$ reaches a peak value rapidly and then develops into a steady value before the cavity pinch-off, which shows good agreement with experimental data. Regarding the pressure at $p2$, without entrained air before cavity pinch-off, the air pressure remains stable and close to $0$.
 
\begin{figure}
   \centering
   \includegraphics[width=0.9\linewidth]{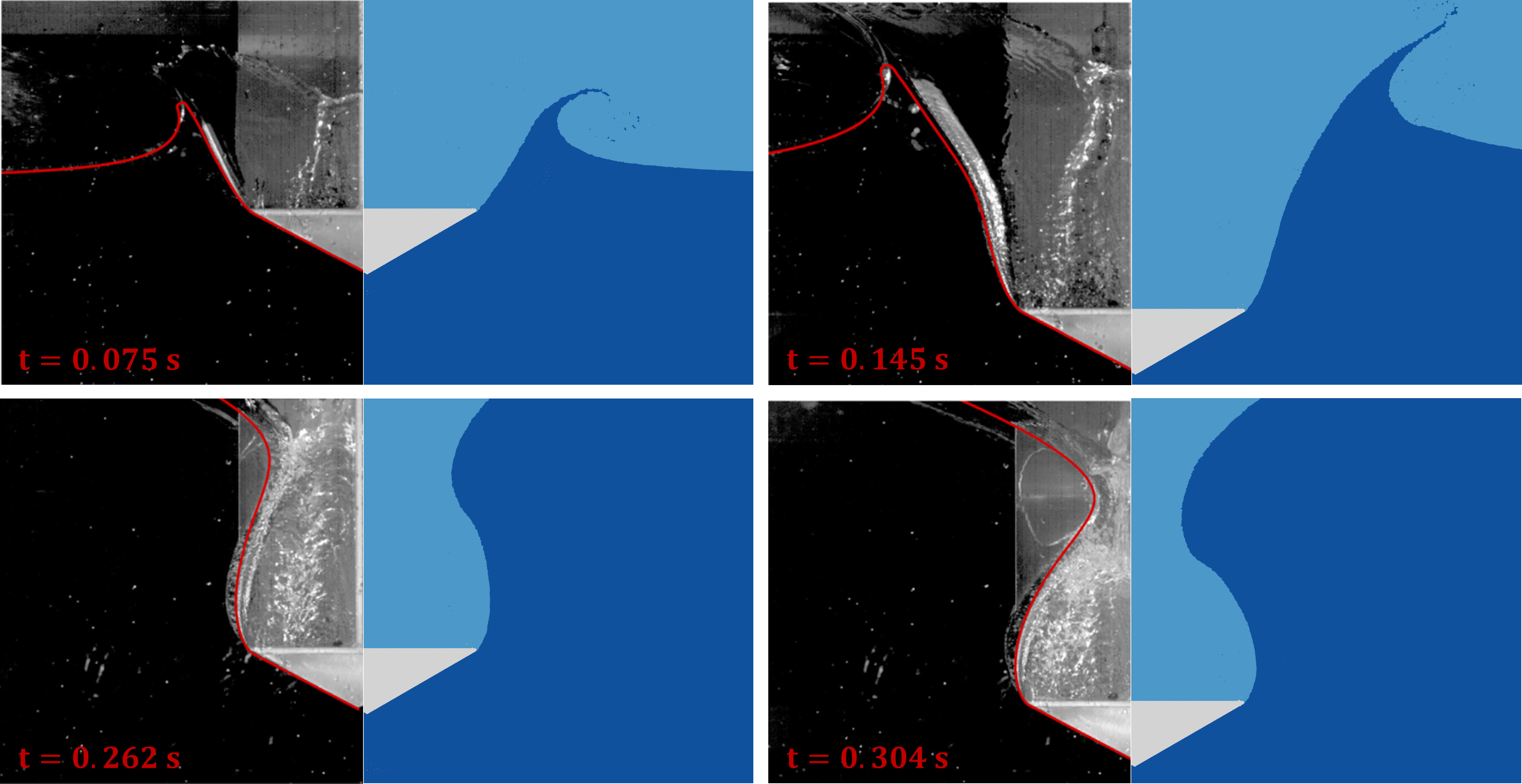}
   \caption{Comparisons between SPH results and experimental snapshots \cite{wang2015experimental} during the water entry stage before the pinching-off of the cavity.} \label{fig:water-entry-snapshot}
 \end{figure}

 Moreover, Figure \ref{fig:water-entry-snapshot} compares the flow simulated by the present multiphase SPH model and obtained by experiment at the same instants before the cavity pinch-off. Good agreement is obtained between them, demonstrating the ability of the present scheme.
 
\begin{figure}
   \centering
   \includegraphics[width=0.95\linewidth]{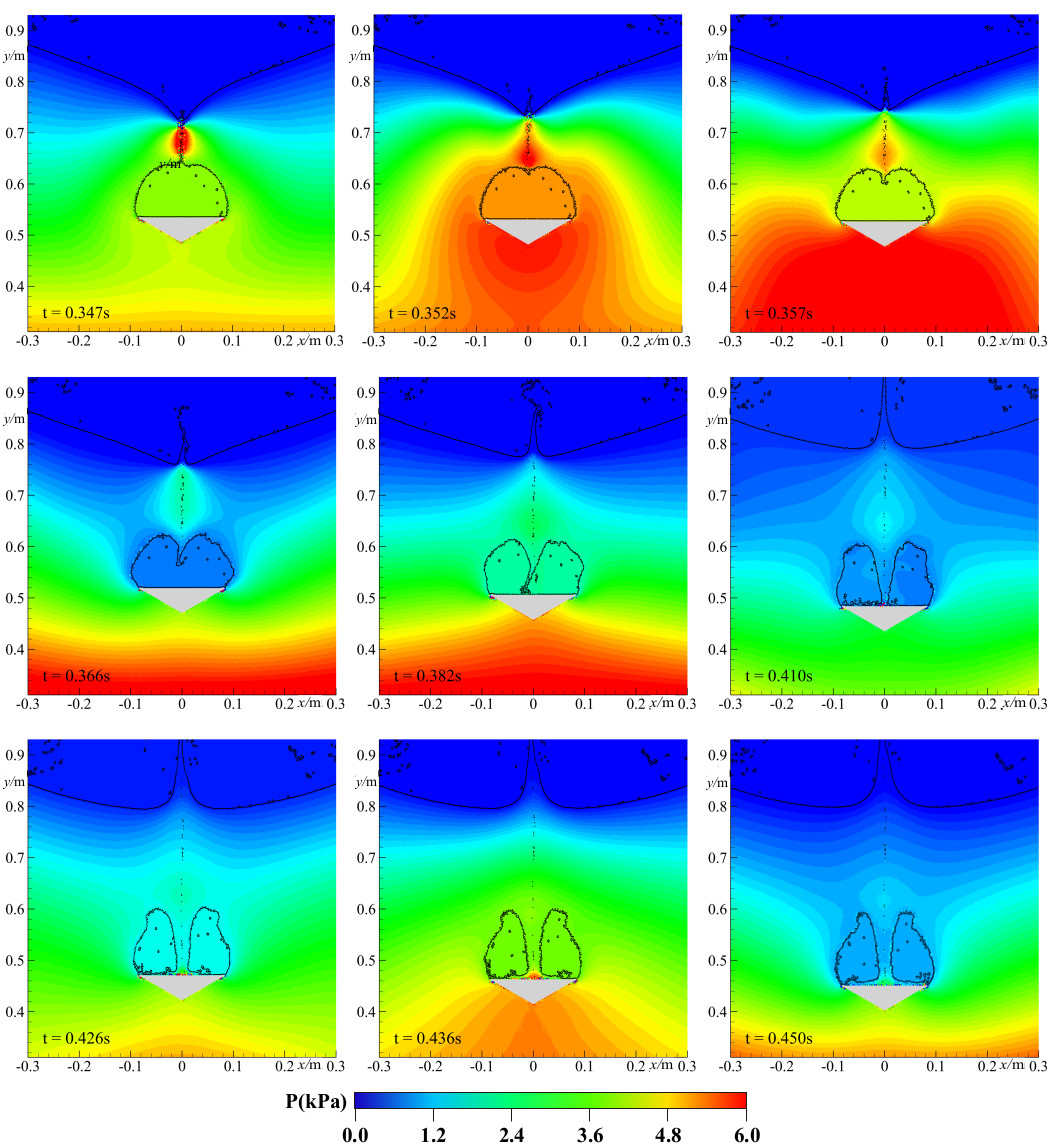}
   \caption{The pressure field during the water entry stage after pinching off of the cavity.}
    \label{fig:water-entry-pinch-off-pressure}
 \end{figure}

 The pressure oscillation in the cavity dominates the water entry dynamics after cavity pinch-off. As depicted in Figure \ref{fig:water-entry-verification}, the oscillation period of the pressure evolutions ($p1$ and $p2$) between the present SPH simulation and experiment \cite{wang2015experimental} is similar, except for the time at pinch-off the open cavity. This slight discrepancy may be attributed to the three-dimensional flow effects. The pressure field just after the cavity pitch-off is depicted in Figure \ref{fig:water-entry-pinch-off-pressure}. At time $t = 0.347$ s, one can observe when the open cavity is about to close, the pressure of the certain points near the pinch-off position is higher than the ambient pressure. At $t = 0.352$ s, a pressure shock wave is generated at the pinch-off position, compressing the cavity with high-pressure water and leading to an increase the pressure of cavity. 
 Importantly, the large-scale pressure wave at the impact position, as discussed in \cite{sun2023inclusion,ju2023study}, is effectively mitigated due to the inclusion of an acoustic damper in the present scheme. As the cavity compresses, the pressure peak reaches the first maximum pressure peak higher than ambient water pressure, causing the cavity to expand and rapidly release pressure, as observed at $t= 0.357$ s and $t= 0.366$ s. Subsequently, as the cavity compresses again, the pressure increases, and the second pressure peak decreases as the result of energy dissipation shown in Figure \ref{fig:water-entry-verification} and $t= 0.382$ s in Figure \ref{fig:water-entry-pinch-off-pressure}. The snapshots of the third pressure oscillation are also depicted in the last row of Figure \ref{fig:water-entry-pinch-off-pressure}.
 
 Based on the above discussion, the present multiphase model can effectively simulate fluid-solid interaction and cavity oscillation dynamics.
 
\subsection{Two-layer liquid sloshing}\label{section:two-layer}
\begin{figure}  
  \centering
  \includegraphics[width=0.85\linewidth]{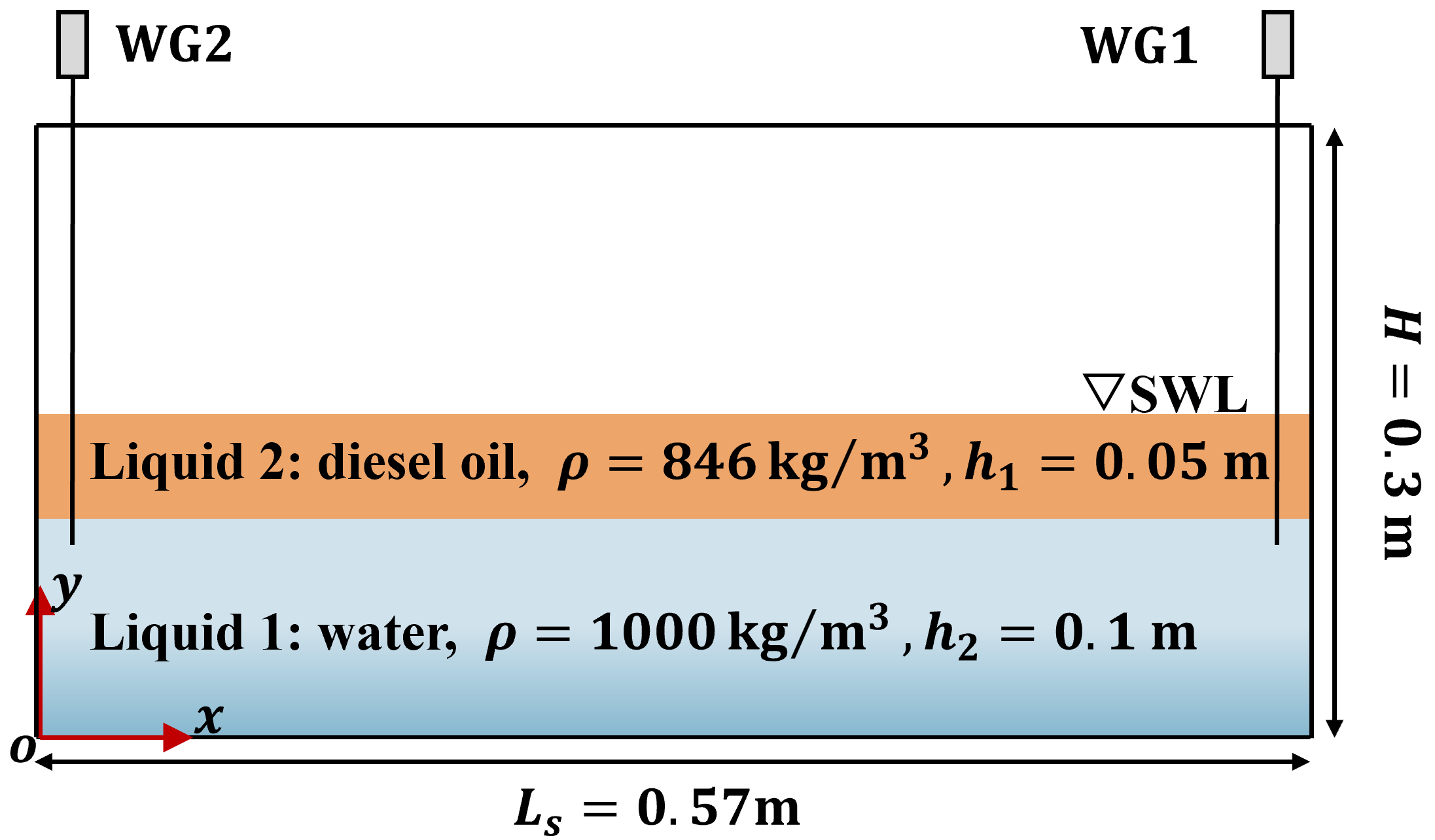}
  \caption{Sketch of the two-layer sloshing case \cite{xue2013experimental}. The rectangular tank with a length of $L_s = 0.57$ m and height of $H = 0.3$ m. } \label{fig:two-layer-sloshing-diagram}
\end{figure}

\begin{figure}
  \centering
  \includegraphics[width=0.9\linewidth]{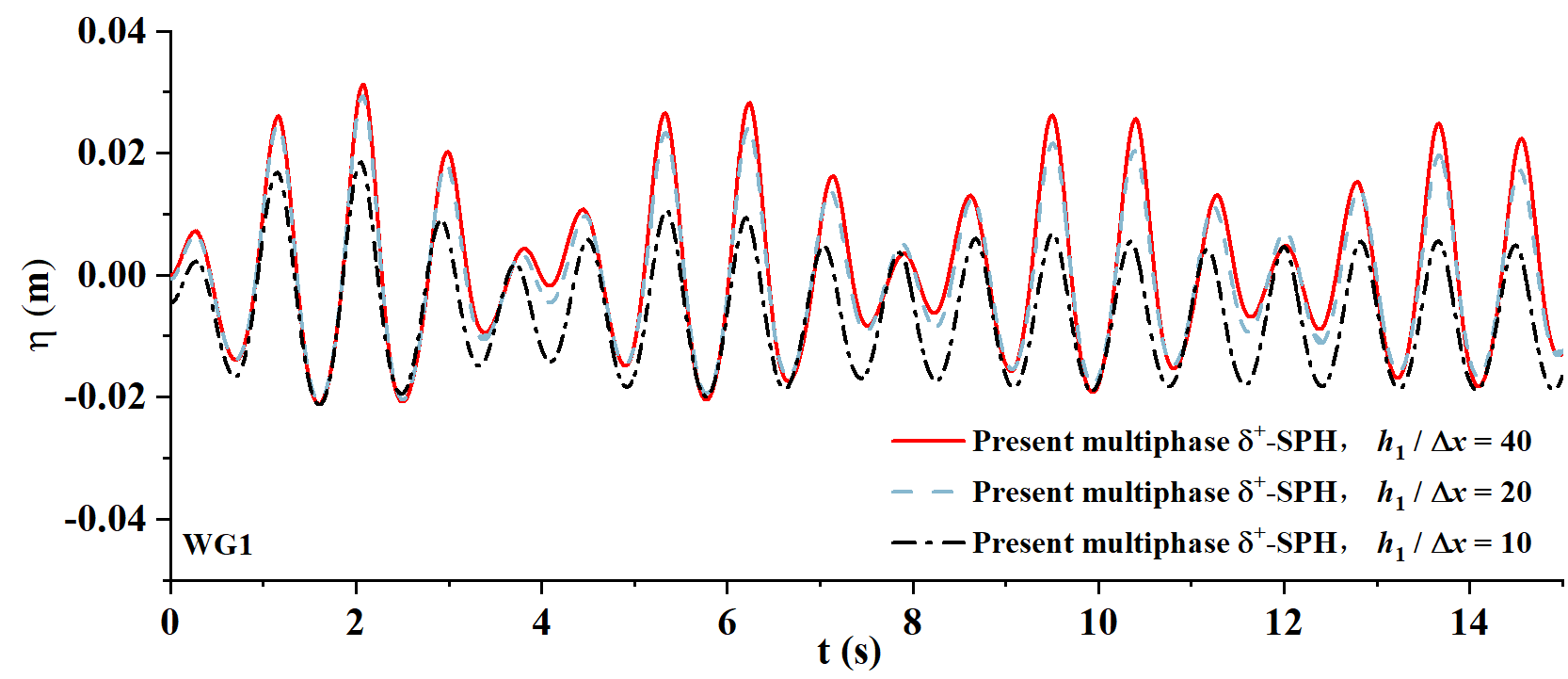}
  \caption{The results of interfacial wave elevation $\eta$ probed at WG1 with particle resolutions $h_1 / \Delta x = 10$, $h_1 / \Delta x = 20$ and $h_1 / \Delta x = 40$.} \label{fig:two-layer-sloshing-convergence}
\end{figure}

\begin{figure}
  \centering
  \includegraphics[width=0.9\linewidth]{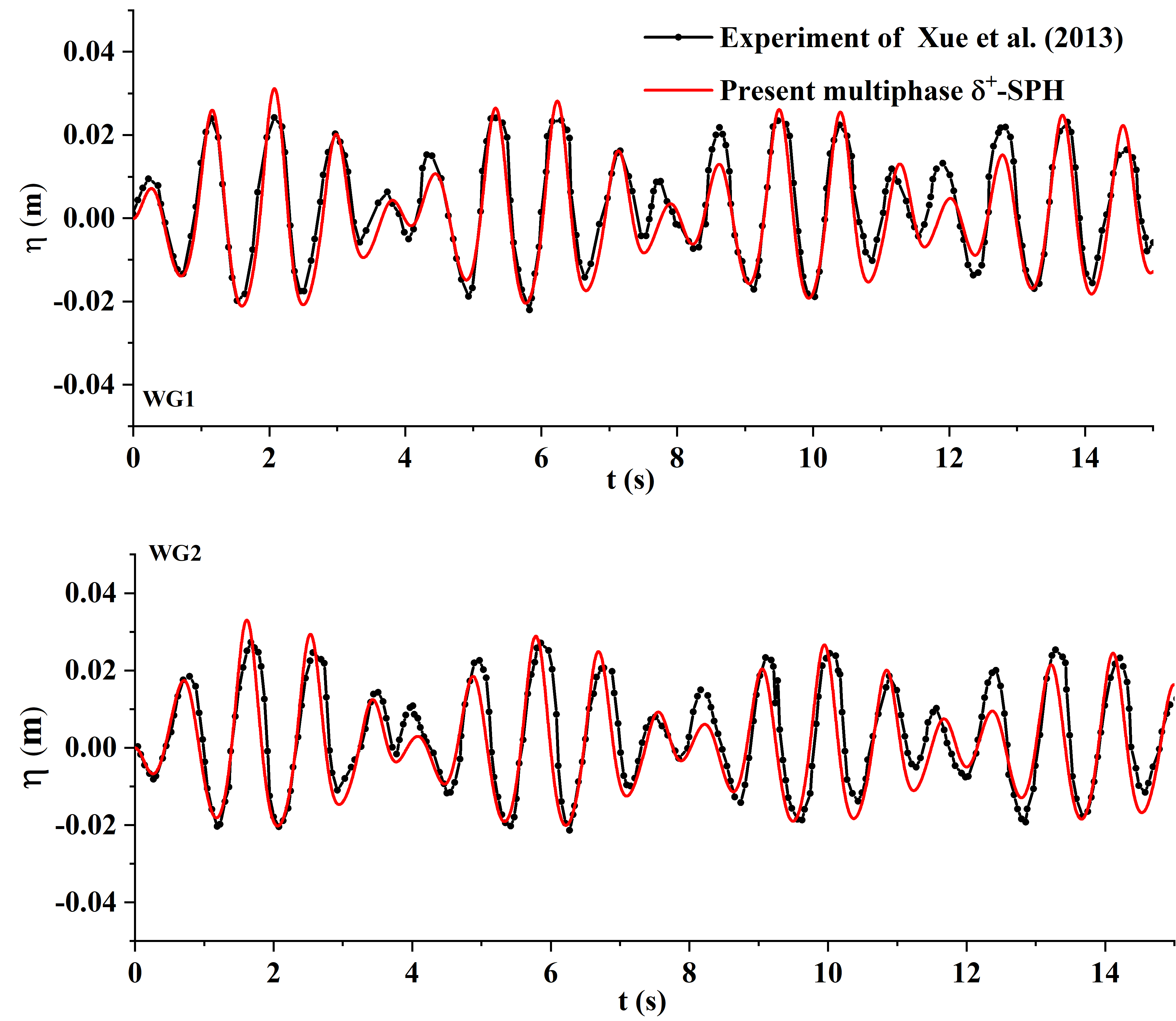}
  \caption{The results of interfacial wave elevation $\eta$ probed at WG1 (top) and WG2 (bottom) compared with experimental data \cite{xue2013experimental}.} \label{fig:two-layer-sloshing-data-verified}
\end{figure}
The sloshing of two-layer case from Xue et al. \cite{xue2013experimental} is simulated to validate the accuracy of the present multiphase $\delta^{+}$-SPH model to handle free-surface problems. The setup is illustrated in Figure \ref{fig:two-layer-sloshing-diagram}. Consistent with the experimental setup, a rectangular tank with a length of $L_s = 0.57$ m is utilized. Its motion is governed by a single-degree-of-freedom motion described as $x(t) = -A\left(\cos\omega t\right)$, where $A = 0.01$ m and $\omega = 7.57$ rad/s. The upper liquid layer is diesel oil with a density of $\rho = 846$ kg/m$^3$ and thickness $h_1 = 0.05$ m, while the lower layer is water with a density of $\rho = 1000$ kg/m$^3$ and thickness $h_2 = 0.1$ m. Two gauges (referred to as WG2 and WG1 in Figure \ref{fig:two-layer-sloshing-diagram}) are placed, respectively, at distances of $10$ cm from the left and right walls to monitor the interfacial wave evaluation $\eta$.

The SPH results for interface wave elevation $\eta$ simulated with three particle resolutions are shown in Figure \ref{fig:two-layer-sloshing-convergence}.
It can be observed that the cycle period and magnitudes of the SPH results with particle resolutions $h_1 / \Delta x = 20$ and $h_1 / \Delta x = 40$ are quite similar.
Therefore, the results with a particle resolution of $h_1 / \Delta x = 40$ show convergence, and this resolution is used for further discussion in this subsection. 
Additionally, the present SPH results for interface wave elevation $\eta$ are verified against the experimental data shown in Figure \ref{fig:two-layer-sloshing-data-verified}, demonstrating good agreement. 

\begin{figure}
  \centering
  \includegraphics[width=0.85\linewidth]{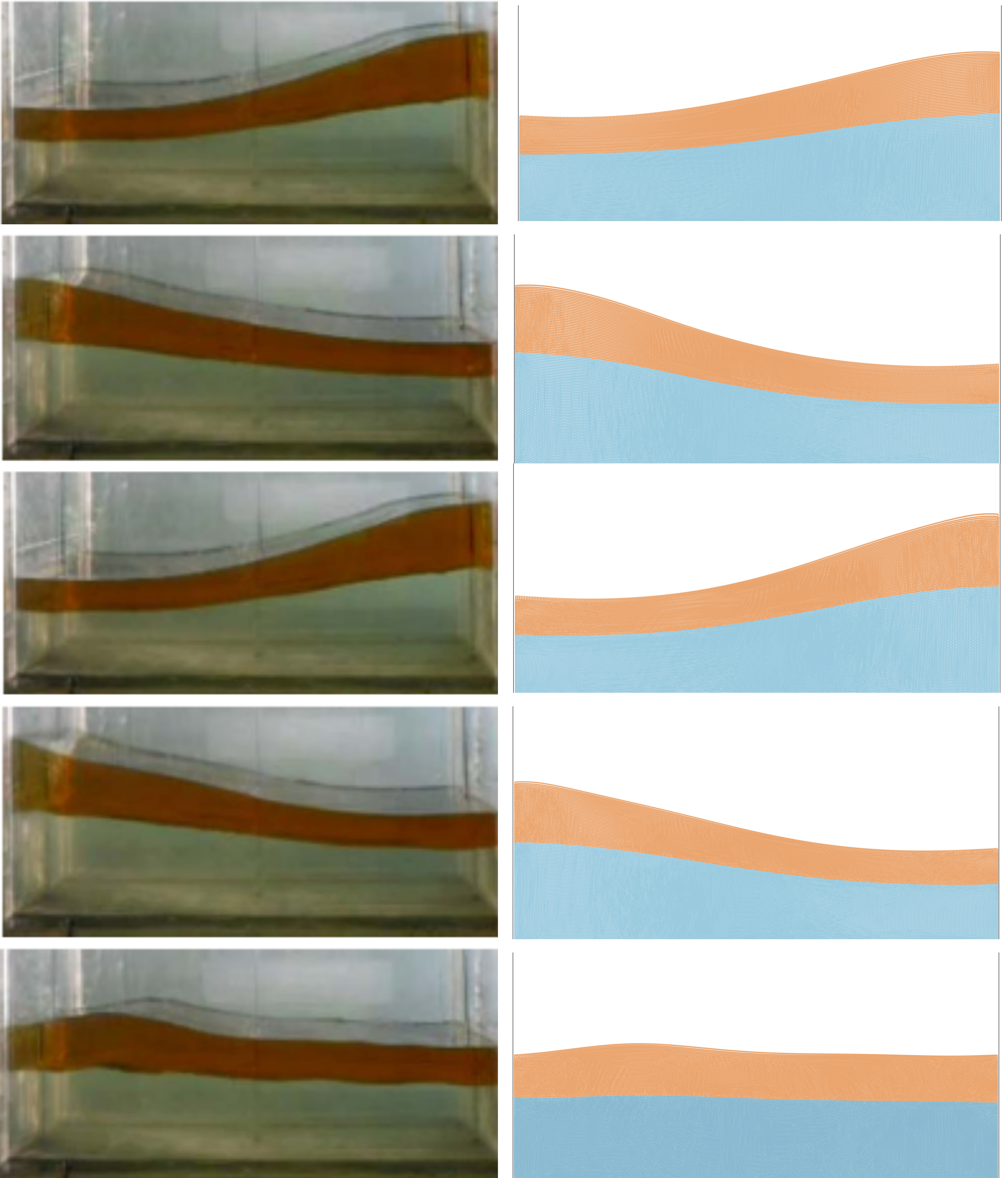}
  \caption{Comparison of the sloshing wave shapes between the results of experimental snapshots \cite{xue2013experimental} (left column) and the present multiphase $\delta^{+}$ SPH model (right column) at $t = 1.20$ $\rm{s}$, $t = 1.65$ $\rm{s}$, $t = 2.10$ $\rm{s}$, $t = 2.55$ $\rm{s}$, $t = 3.25$ $\rm{s}$, respectively, from top to bottom.} \label{fig:two-layer-sloshing-snapshots}
\end{figure}

Figure \ref{fig:two-layer-sloshing-snapshots} presents a comparison between the present SPH results and the experimental snapshots at different times. The SPH results including the free surface and multiphase interface, show good agreement with the experimental snapshots. 
A smooth and accurate interface is obtained in the present work, demonstrating the capability of the current multiphase SPH model to simulate multilayer sloshing problems effectively. 

\begin{figure}
  \centering
  \subfigure{
    \includegraphics[width=0.9500\linewidth]{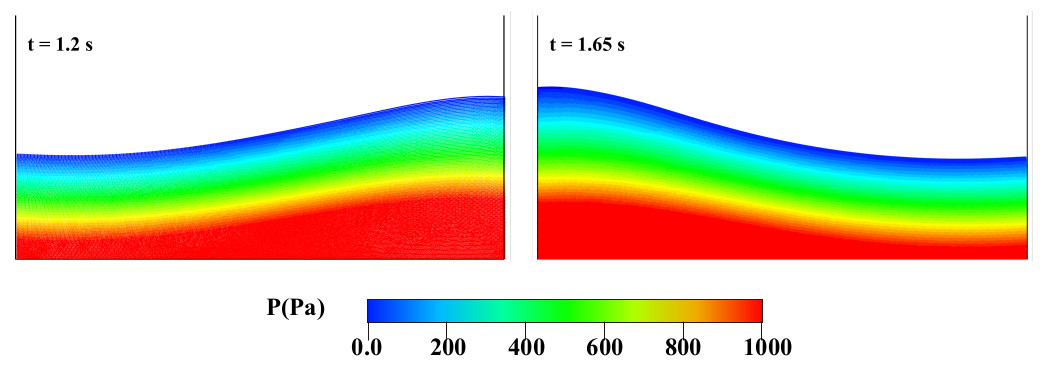}
  }
  \caption{The pressure field of two-layer liquid sloshing at $t = 1.2, 1.65$ \rm{s}. }
  \label{fig:two-layer-sloshing-pressure}
\end{figure}

Taking the instants $t = 1.2$ s and $t = 1.65$ s when the free surface exhibits large deformation as examples, as shown in Figure \ref{fig:two-layer-sloshing-pressure}, the pressure field is also smoothed and without noise at the interface. Therefore, it is sufficient to demonstrate that the present SPH scheme has a good capability to handle free-surface problems. 

\begin{figure}
  \centering
  \subfigure{
    \includegraphics[width=0.98\linewidth]{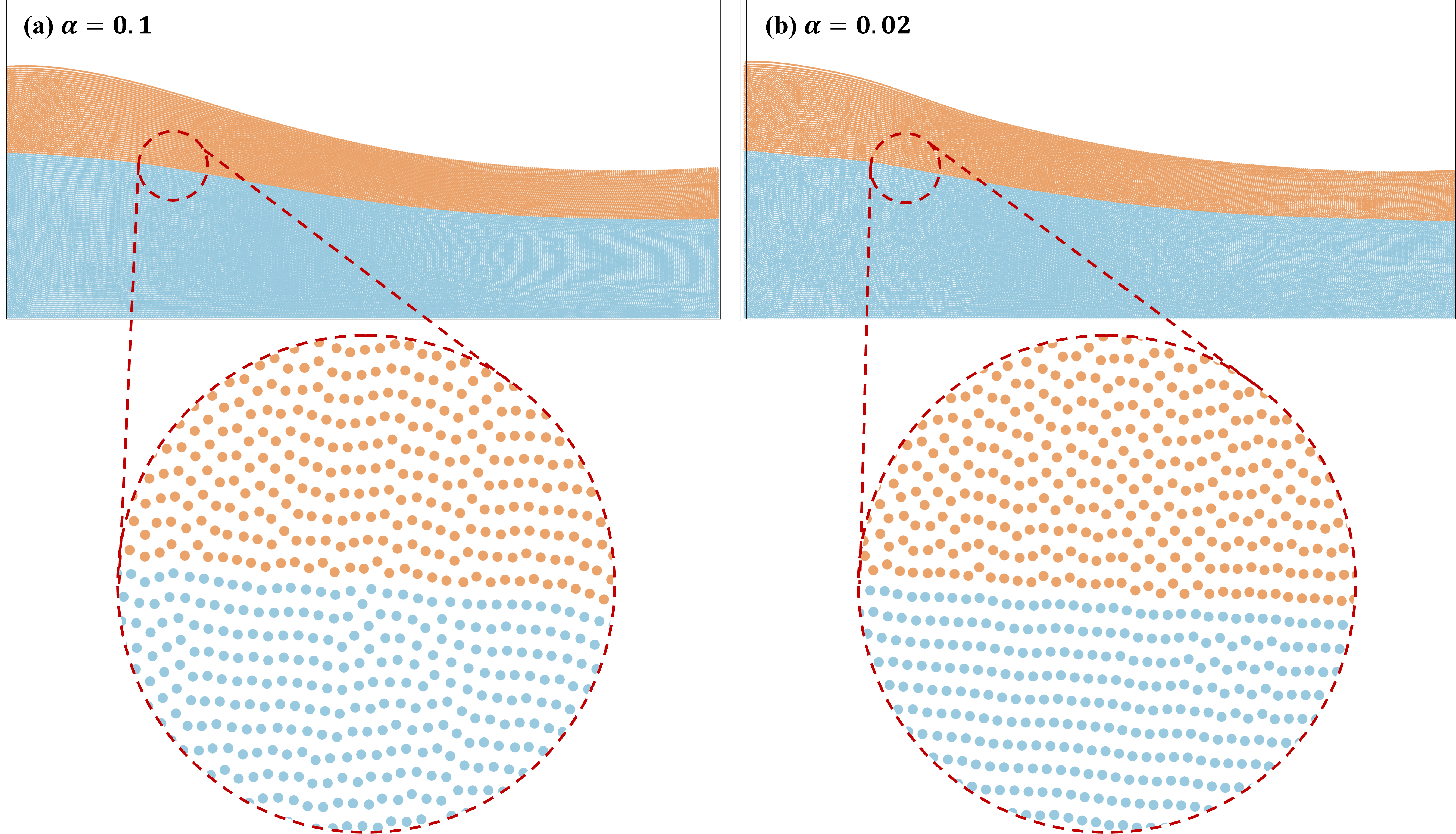}
  }
  \caption{The particle distribution of two-layer liquid sloshing at $t = 1.65$ \rm{s}, simulated with $\alpha =0.1$ and $\alpha =0.02$, respectively. }
  \label{fig:two-layer-sloshing-particle-alpha}
\end{figure}

\textcolor{black}{Additionally, we also compare the particle distribution with the results simulated with the artificial viscosity parameter $\alpha = 0.02$, as shown in Figure \ref{fig:two-layer-sloshing-particle-alpha}. The interface and free surface simulated with $\alpha = 0.02$ are similar to those simulated with $\alpha = 0.1$, and the particle distribution is uniform with no penetration occurring. It is again indicated that artificial viscosity does not play a key role in preventing interface penetration.}

\subsection{Water sloshing with entrapped air pockets}\label{air-pockets}
\begin{figure}
  \centering
  \includegraphics[width=0.85\linewidth]{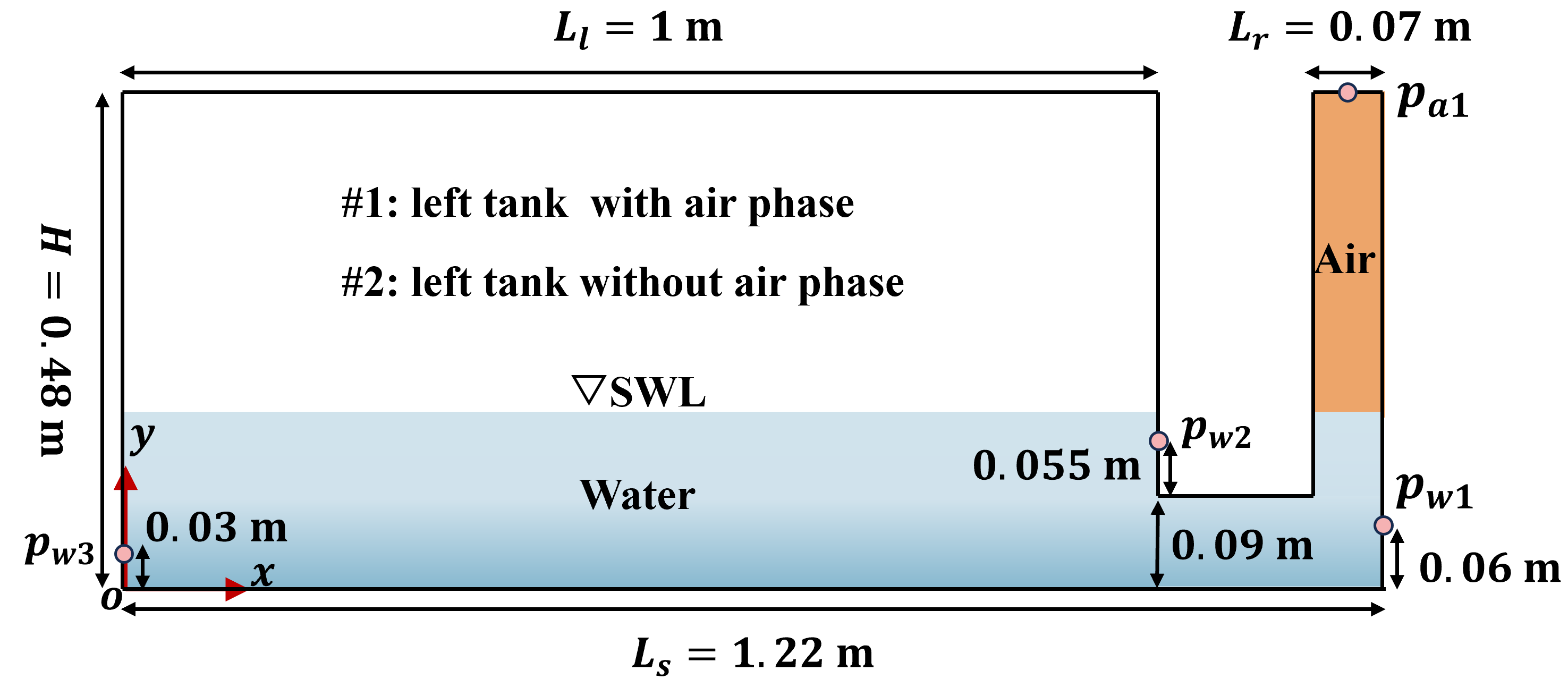}
  \caption{
    A schematic of the tank simulated with water sloshing with an entrapped air pocket \cite{luo2016particle}. Two cases are considered: one with the remaining region filled with air phase (referred to as $\#1$), and the other without (referred to as $\#2$) in the left tank.} \label{fig:air-pocket-diagram}
\end{figure}

The air pressure predicted in the investigations is limited in the previous literature. In this subsection, the water sloshing with an entrapped air pocket, experimentally investigated by Luo et al. \cite{luo2016particle} is simulated in the present work to verify the accurate prediction of air pressure. 
The numerical tank setup is identical to the experiment \cite{luo2016particle} shown in Fig. \ref{fig:air-pocket-diagram}. The tank is divided into left and right tanks connected by a short channel, each with a water depth of $d_l = d_r = 0.17$ $\rm{m}$. The motion of the tank follows a traditional pattern \cite{luo2016particle} governed by:
\begin{equation}
  x\left(t\right)=
  \begin{dcases}
   - \frac{At}{5} \sin \left(\omega t\right)\,,\,\,\,\,\,\,\,\, t<5 \,\rm{s} \\
   - A \sin \left(\omega t\right)\,,\,\,\,\,\,\,\,\,   t\geq 5 \,\rm{s},
  \end{dcases}
\end{equation}
where $A$ and $\omega$ are the amplitude and frequency, set as $A = 0.0412$ $\rm{m}$, and $\omega = 3.6807$ $\rm{rad/s}$. In this configuration, the air phase is compressed in the right tank when water sloshes from left to right, and is expanded when sloshing from right to left inducing significant variation of air pressure. Four pressure sensors are placed: $p_{w1}$ and $p_{a1}$ placed at the right tank to monitor the water and air pressure, respectively, and $p_{w2}$  and $p_{w3}$ to monitor the water pressure in the left tank.
Notably, the left tank is connected to the atmosphere and no air is trapped in it, allowing the air phase in the remaining region of the left tank to be neglected. In this scenario, the capability of the present scheme to model multiphase flows with free-surface can be more clearly observed. Additionally, as mentioned in subsection \ref{section:dam-breaking}, the present work can maintain stable pressure of the air phase with the ghost boundary condition. Therefore, to further describe the pressure variation of the air phase, the case of retaining the air phase in the left tank (i.e. $\#1$ shown in Figure \ref{fig:air-pocket-diagram}) is also simulated for comparison with the former setup. The fluid properties of water and air are consistent with those in the previous cases.

\begin{figure}
  \centering
  \includegraphics[width=0.95\linewidth]{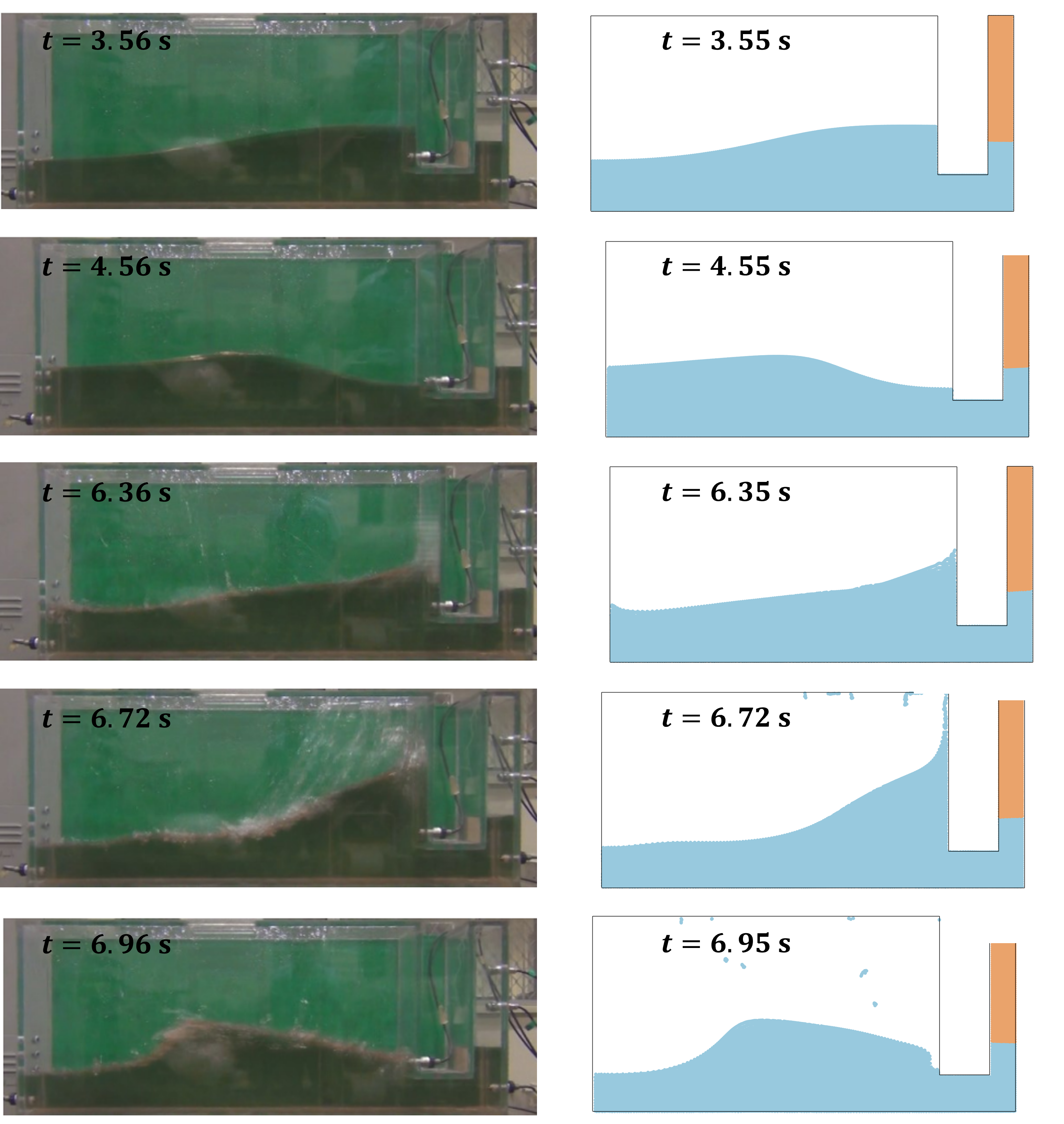}
  \caption{Comparison of the sloshing wave shapes between the results of the consistent multiphase $\delta^{+}$-SPH model (right column) and experiment (left column) \cite{luo2016particle}. } \label{fig:air-pocket-snapshot}
\end{figure}

As shown in Figure \ref{fig:air-pocket-snapshot}, the wave shape simulating with the present SPH model shows ($\#2$ in this case) relatively good agreement with the experimental wave profiles at several time instants. As can be observed, smoothed free surface is reproduced without additional treatment, illustrating the capability to accurately handle free-surface in the present model. 

\begin{figure}
  \centering
  \subfigure[The evolution of pressure probed at point $p_{w2}$]{
    \includegraphics[width=0.950\linewidth]{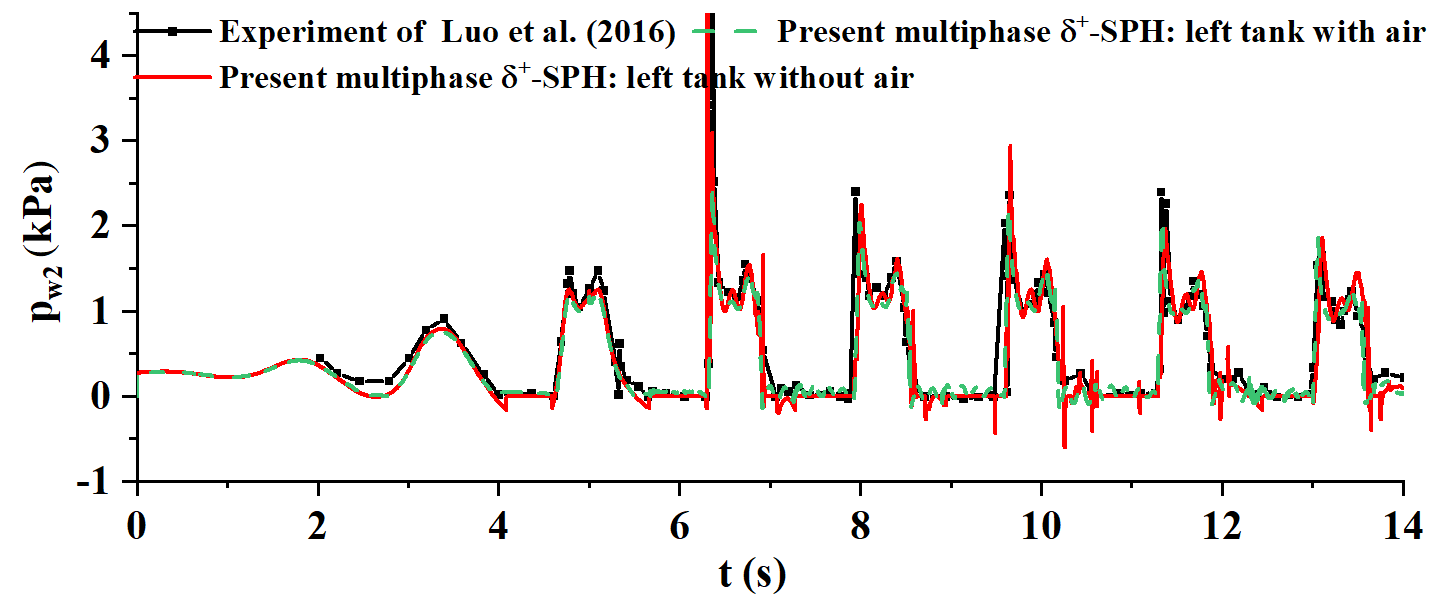}
  }
  \subfigure[The evolution of pressure probed at point $p_{w3}$]{
    \includegraphics[width=0.950\linewidth]{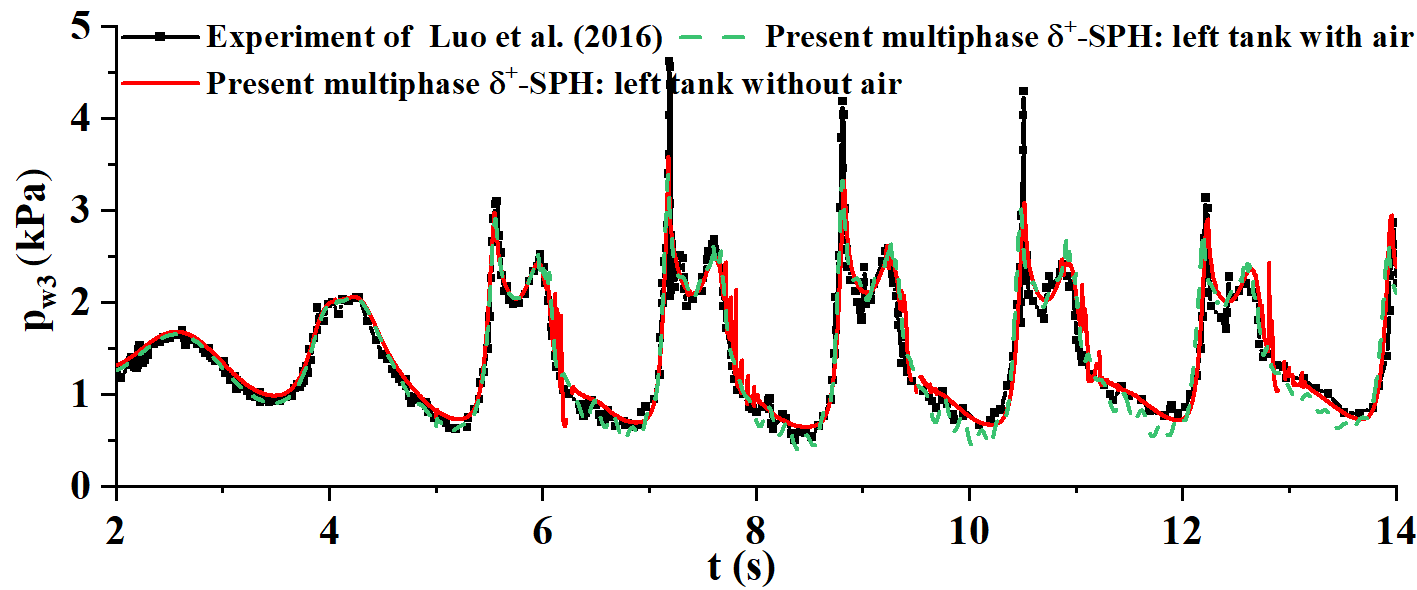}
  }
  \caption{The present SPH results including $\#1$, and $\#2$ scenarios of water pressure monitored at points $p_{w2}$ and $p_{w3}$ compared with experimental results \cite{luo2016particle}.}
  \label{fig:air-pocker-left-tank-pressure}
\end{figure}

 Figure \ref{fig:air-pocker-left-tank-pressure} presents the water pressure monitored at $p_{w2}$ and $p_{w3}$. The impact pressures on the solid wall simulating with the present multiphase $\delta^{+}$-SPH model agree well with the experimental data \cite{luo2016particle}. 

\begin{figure}
  \centering
  \includegraphics[width=0.95\linewidth]{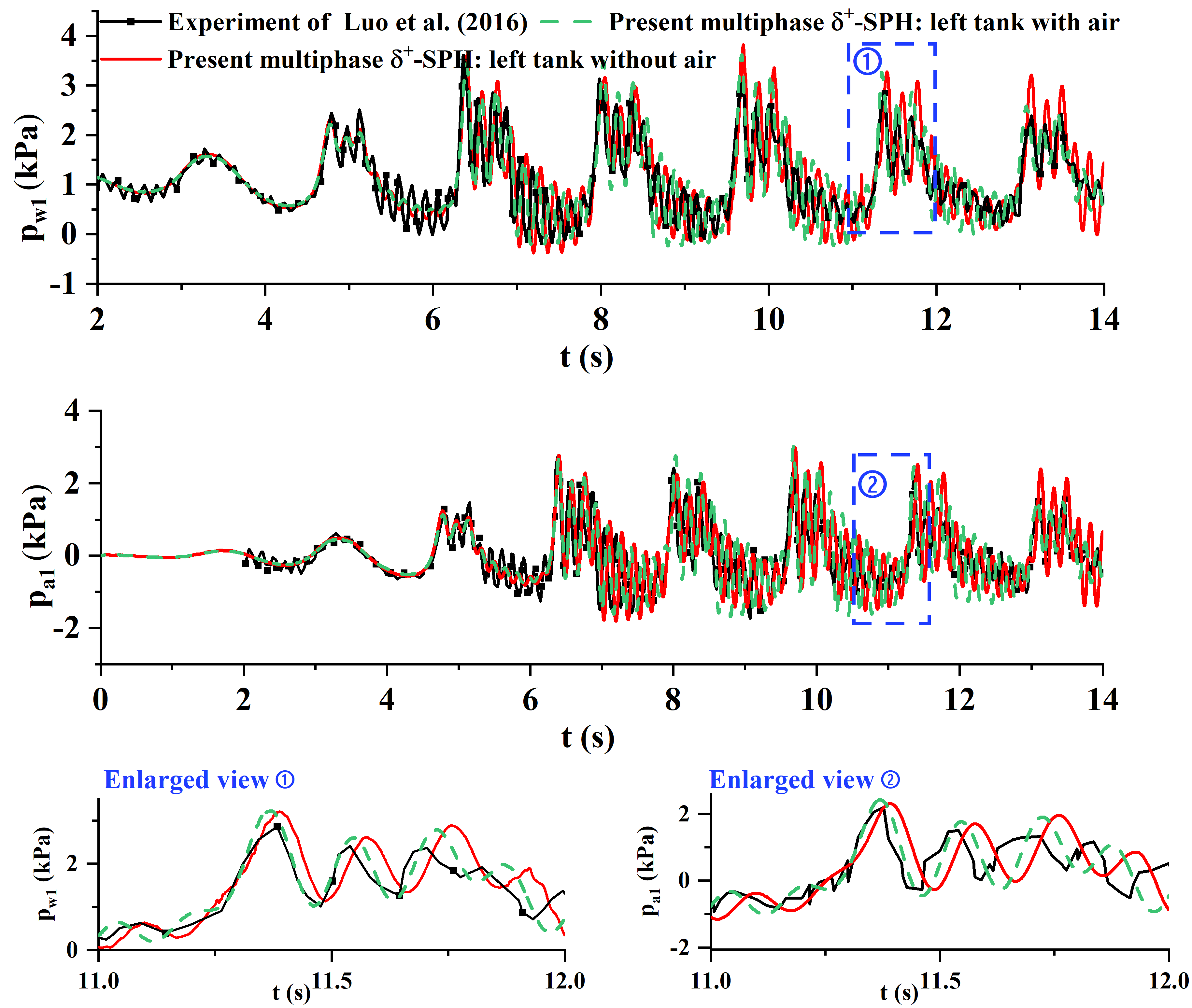}
  \caption{The present SPH results (with and without air phase in the left tank) of air pressure monitored at $p_{a1}$ and water pressure at $p_{w1}$ compared with experimental results \cite{luo2016particle}.} \label{fig:air-pocket-right-tank-pressure}
\end{figure}

\begin{figure}
  \centering
  \includegraphics[width=0.95\linewidth]{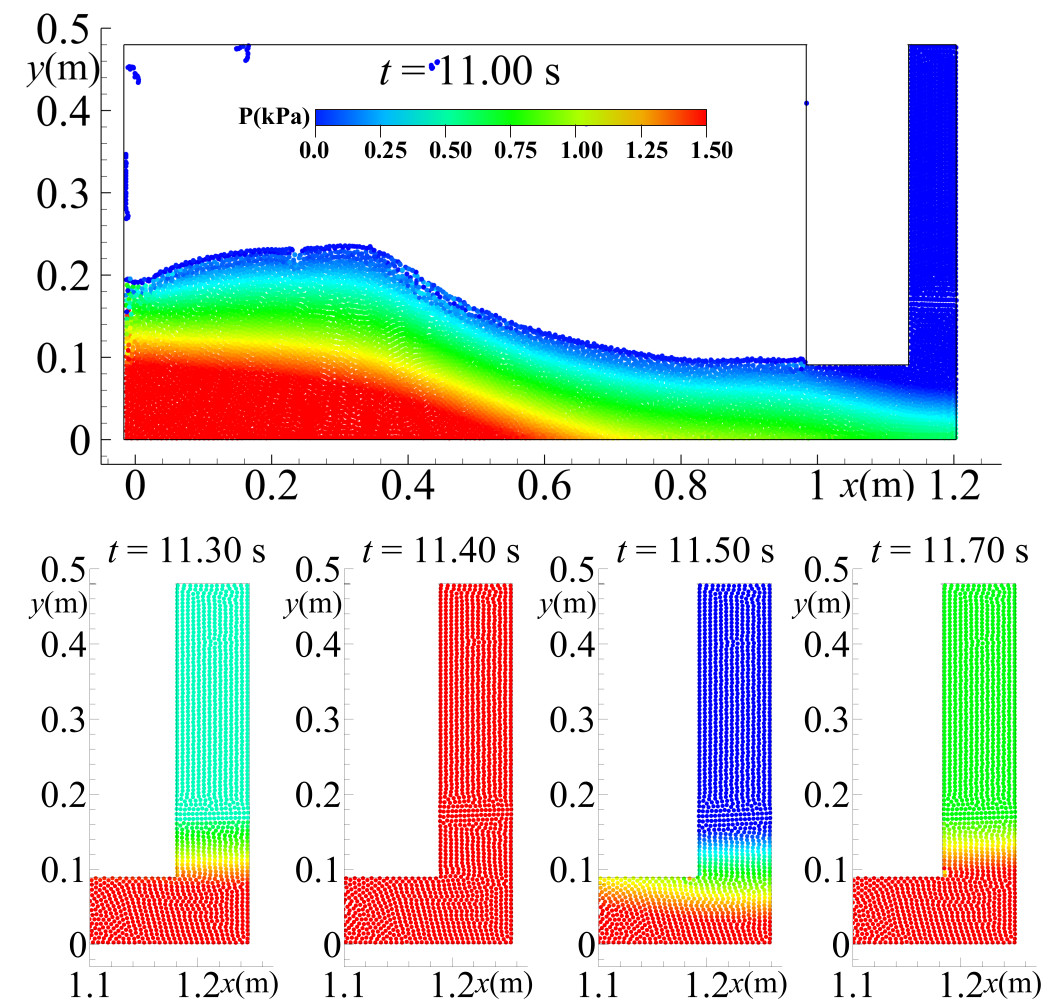}
  \caption{The pressure fields of right tank at several instants form$ t = 11 s$ to $t = 12 s$.} \label{fig:fig:air-pocket-right-tank-contour}
\end{figure}

Additionally, Figure \ref{fig:air-pocket-right-tank-pressure} depicts the temporal evolution of the water and gas pressures monitored at the right tank. It is noteworthy that both the SPH simulation results, with or without the air phase in the left tank, show good agreement with the experimental data. Upon impact on the water in the connecting channel, resulting in a rise in pressure, the air pocket compresses as expected (see the pressure point $p_{a1}$). Subsequently, the air pressure influences the water pressure surrounding the air pocket as well (see the pressure point $p_{w1}$). Correspondingly, as shown in Figure \ref{fig:air-pocket-right-tank-pressure}, the trend of the air and water pressure variation observed in the right tank closely resembles each other.
As plotted in the last row, during the sloshing process from $t = 11$ $\rm{s}$ to $t = 12$ $\rm{s}$, the enlarged views of pressure variation of both air and water show good agreement despite some discrepancies. Furthermore, the evolution of air pressure contours during time $t = 11$ $\rm{s}$ to $t = 12$ $\rm{s}$ is plotted in Figure \ref{fig:fig:air-pocket-right-tank-contour}. As can be seen, the pressure of the air and the interphase are smoothed despite large density ratio.

In general, as shown in Figure \ref{fig:air-pocket-right-tank-pressure} and Figure \ref{fig:fig:air-pocket-right-tank-contour} it is sufficient to demonstrate the strong capability of the present SPH scheme to model the compressibility of air and predict real air pressure variation in multiphase flow problems.

\begin{figure}
  \centering
  \includegraphics[width=0.95\linewidth]{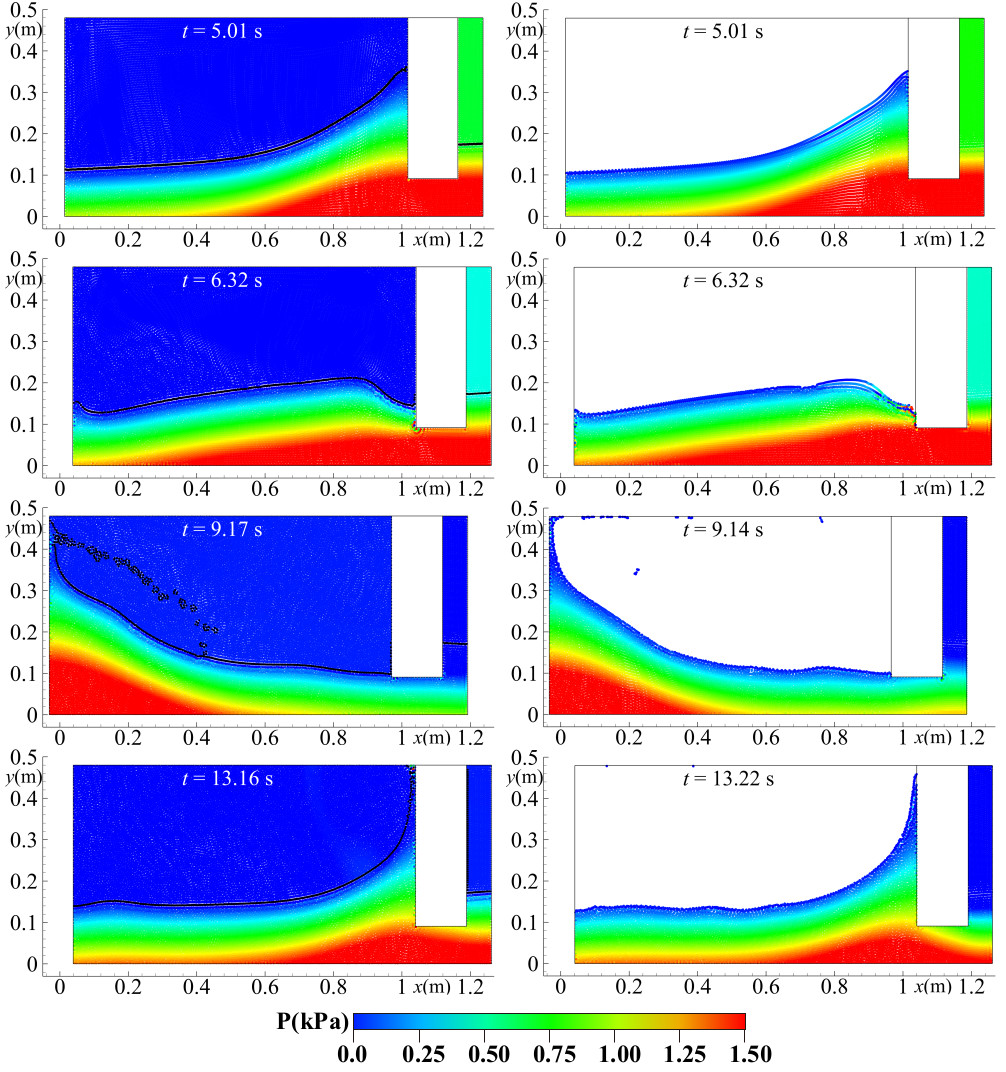}
  \label{fig:1}
  \caption{Comparison of pressure fields between simulations with (left column) and without (right column) the air phase in the left tank. } \label{fig:with-with-air-pressure}
\end{figure}

Figure \ref{fig:with-with-air-pressure} presents a comparison of pressure fields between simulations with and without an air phase in the left tank, representing scenarios with and without a free surface, respectively. Notably, the wave profiles simulated with a free surface are closely consistent with those of the scenario without it. Since there is no air pocket in the left tank, water pressure fields are also consistent with each other. This is evident from the results shown in Figures \ref{fig:air-pocker-left-tank-pressure} and \ref{fig:air-pocket-right-tank-pressure}, fully demonstrating the accuracy and capability of the consistent $\delta^{+}$-SPH in handling free surface.

\section{Conclusion}\label{conclusion}

In this study, we introduce a consistent $\delta^{+}$-SPH model for multiphase flows, expanding upon the framework outlined for single-phase flows in \cite{sun2019consistent}. The consistent multiphase $\delta^{+}$-SPH model accounts for the compressibility of different phases, integrating an acoustic damper term for incompressible phases and using physical sound speeds for compressible phases.
Distinguished from the work by Sun et al. \cite{sun2019consistent} on single-phase flows, the proper implementation of the shifting velocity at interfaces poses a unique challenge for multiphase problems. To address this issue, we proposed new strategies for handling the $\delta \boldsymbol{u}$ terms at multiphase interfaces, ensuring stable pressure fields for both water and air without requiring a background pressure.

We conducted a two-phase hydrostatic test against analytical trends, demonstrating the stability and accuracy of the current multiphase model.
A smoothed interface, as well as the particle distribution and pressure field, was obtained, with the pressure profiles agreeing well with analytical results.

For what concerns impact flows with entrapped air problems, multiphase dam breaking, slamming of LNG tank insulation panel, the cavity oscillation during wedge water entry, and sloshing with entrapped air pocket test cases are considered all characterized by large density ratios. 
In the dam breaking case, unlike the $\delta^{+}$-SPH model with simple particle shifting \cite{sun2017deltaplus}, the present multiphase $\delta^{+}$-SPH model effectively accounted for air compressibility and achieved stable pressure field, showing a better performance in volume conservation. In the other three cases, the air pressure was qualitatively verified, and in particular, pressure fluctuations over short durations were accurately predicted by monitoring the pressure of the entrapped air pocket or air cavity. Furthermore, to demonstrate the capability of the present SPH model in modeling free surfaces in multiphase flows, we conducted two-layer liquid and sloshing with an entrapped air pocket under two conditions: with and without free-surface in the left tank. Both wave profiles and pressure loads on the tank were consistent, showing the capability to solve with free-surface in multiphase scheme.

The good capability of the present SPH model in modeling the compressibility of different phases was well discussed. We expect to extend the present multiphase SPH scheme to applications involving violent flows with entrained air, such as violent wave impact on structures.

\section*{Acknowledgment}
This research was partially funded by the National Natural Science Foundation of China (Grant No. \textcolor{blue}{52171329}), the Guangdong Basic and Applied Basic Research Foundation (Grant No. \textcolor{blue}{2024B1515020107}) and Characteristic Innovation Project of Universities in Guangdong Province (Grant No. \textcolor{blue}{2023KTSCX005}). 
The research leading to these results was partially funded by the HASTA project (Grant No. \textcolor{blue}{101138003}) as part of the European Union Horizon research programme. Views and opinions expressed are however those of the authors only and do not necessarily reflect those of the European Union. Neither the European Union nor the granting authority can be held responsible for them.
We would like to express our sincere gratitude to the reviewers for their detailed review and valuable feedback.
\section*{  }
\appendix
 \section{\textcolor{black}{Enumeration tests: two strategies for the treatment of \texorpdfstring{$\delta \bm{u}$}--terms in continuity equation applied at interfaces}} \label{section:appendix}

 \textcolor{black}{We proposed two strategies for handling $\delta \bm{u}$-terms in the continuity equation, specifically, the strategy $\divideontimes 1$ refers to Eq. \ref{eq:swith-for-fluid-solidinterface} and the strategy $\divideontimes 2$ refers to Eq. \ref{eq:multi-fluid-interface} for $\delta \bm{u}$-terms in the continuity equation as discussed in Subsection \ref{section:interface-treatment}. Here, through cases of two-phase hydrostatic and dam-break scenarios, we evaluate the performance of the two strategies at interfaces, including fluid-solid and multi-fluid interfaces. 
Notably, in the present SPH, strategy $\divideontimes 1$ is applied at fluid-solid interfaces and the strategy $\divideontimes 2$ at multi-fluid interfaces, showing stable behavior (labeled as \smiley). Four modes discussed for $\delta \bm{u}$-terms treatment at interfaces are listed in Table \ref{tab:four_mode_for_delta_term}.
 } 
\begin{table}[H]
  \centering
  \caption{Modes of different strategies for handling $\delta \bm{u}$-terms in the continuity equation at the interfaces. Strategy $\divideontimes 1$ refers to Eq. \ref{eq:swith-for-fluid-solidinterface} and strategy $\divideontimes 2$ refers to Eq. \ref{eq:multi-fluid-interface}.}\label{tab:four_mode_for_delta_term}
  \begin{tabular}{ccccc}
    \toprule
    &Mode & Fluid-solid interface & Multi-fluid interface &   \\    
    \midrule%
    & 1        & $\divideontimes 1 $ \smiley  &  $ \divideontimes$ 1 \frownie &\\    
    & 2        & $\divideontimes 2 $ \frownie  &  $ \divideontimes$ 2 \smiley & \\   
    & 3        & $\divideontimes 2 $ \frownie  &  $ \divideontimes$ 1  \frownie &    \\
    & 4 (Present SPH) & $\divideontimes 1$ \smiley & $ \divideontimes$ 2 \smiley & \\                                      
    \bottomrule
  \end{tabular}
\end{table}
\begin{figure}
  \centering
  \includegraphics[width=0.65\linewidth]{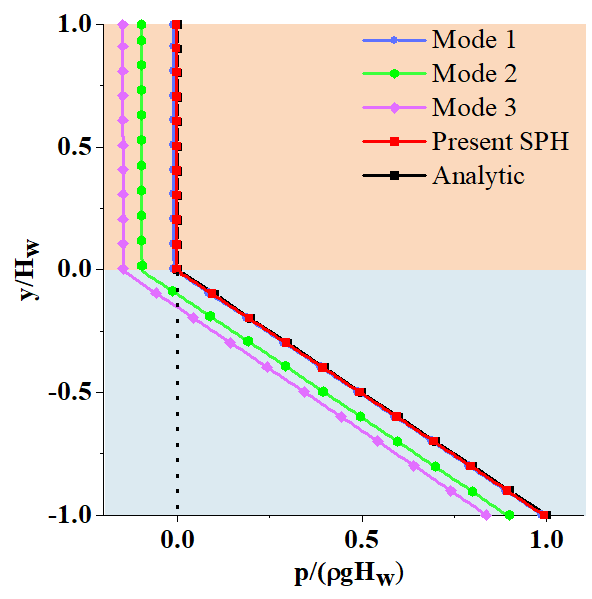}
 
  \caption{The pressure distribution of two-phase hydrostatic at $t\left(g/H_w\right)^{1/2}=40$ simulated with four modes respectively.} \label{fig:hy-multi-fluid-interface-discussion} 
\end{figure}
\textbullet{} \textcolor{black}{\textbf{Comparison of two strategies applied at multi-fluid interface}}

\textcolor{black}{
We first compare the two strategies applied at the multi-fluid interface, with the strategy $\divideontimes 1$ used at the fluid-solid interface. In other words, the comparison is between Mode 1 and Mode 4 (i.e. present SPH). 
}

\begin{figure}
  \centering
  \includegraphics[width=0.75\linewidth]{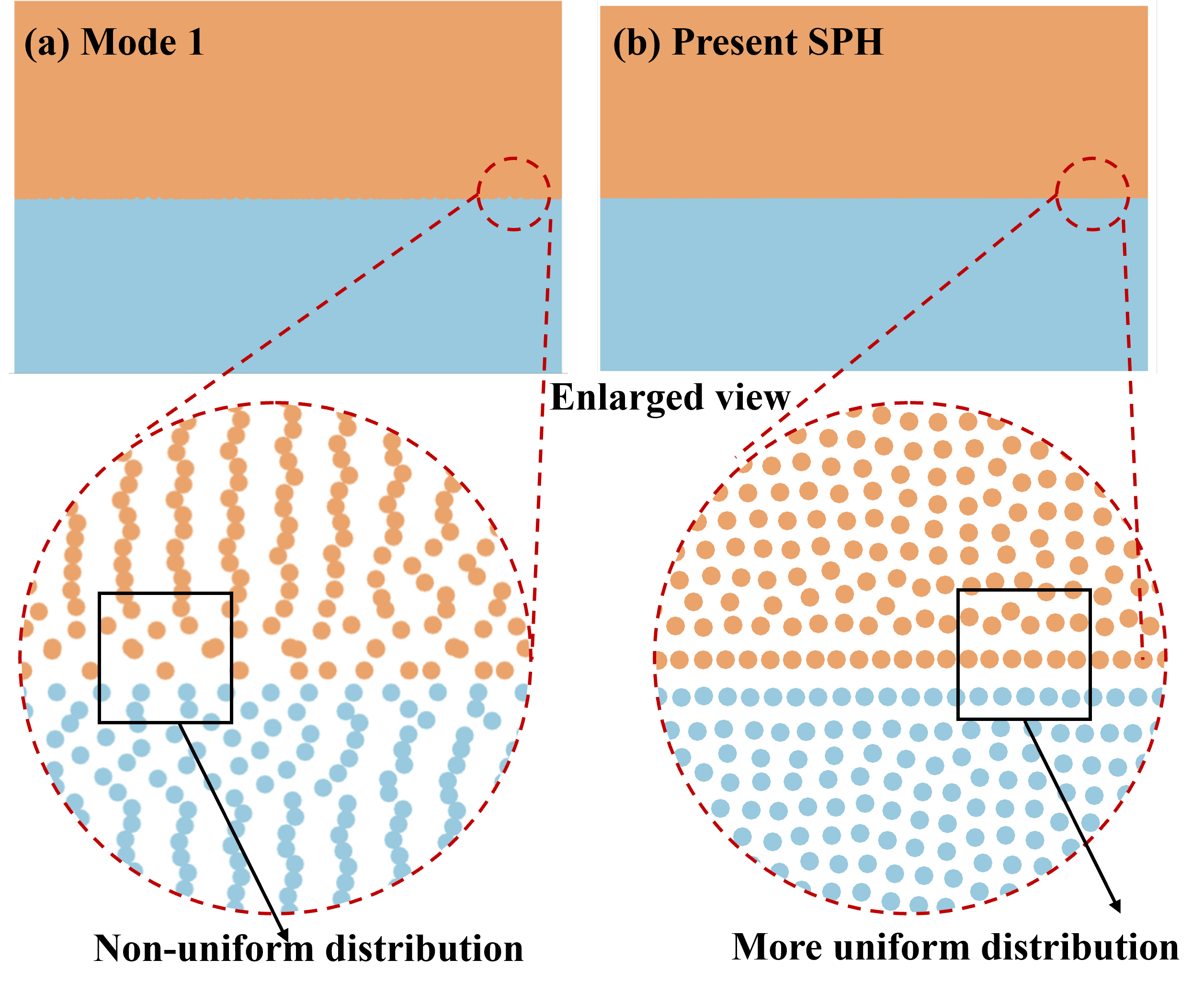}
  \caption{The particle distribution of two-phase hydrostatic simulated with Mode 1 and present multiphase $\delta^{+}$-SPH.}\label{fig:hy-multi-fluid-interface-particle-ditribution-discussion} 
\end{figure}

\textcolor{black}{
In hydrostatic scenarios, Figure \ref{fig:hy-multi-fluid-interface-discussion} shows the pressure distribution along the tank centerline at $t\left(g/H_w\right)^{1/2} = 40$. Both Mode 1 and Mode 4 (that is, the current SPH adopted) achieve accurate pressure distribution and maintain the background pressure. Strategy $\divideontimes$ 1 may be a potential alternative for the multi-fluid interface.
However, as shown in Figure \ref{fig:hy-multi-fluid-interface-particle-ditribution-discussion}, results with Mode 1 show particle clustering at the interfaces, while the present SPH (using strategy $\divideontimes$ 2 at multi-fluid interfaces) results in more uniform particle distribution and stable background pressure.
}

\begin{figure}
  \centering
  \includegraphics[width=0.98\linewidth]{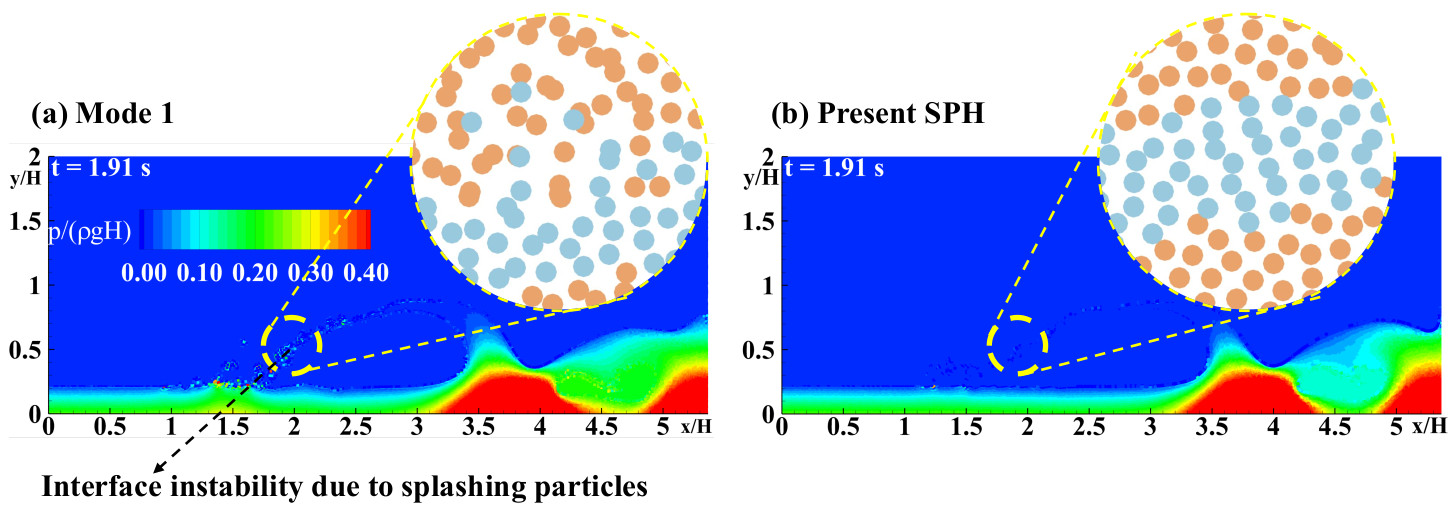}
  \label{fig:1}
  \caption{Comparison of pressure fields and particle distribution (enlarged view) at multi-fluid interfaces simulated with Mode 1 and the present multiphase $\delta^{+}$-SPH.} \label{fig:dambreaking-14-mode}
\end{figure}

\textcolor{black}{In cases of violent impacts, the use of strategy $\divideontimes 1$ exhibits unstable behavior at multi-fluid interfaces, especially instability caused by splashing of fluid particles during large interface deformation. Using the dam-breaking benchmark as an example,
Figure \ref{fig:dambreaking-14-mode} compares particle distributions between Mode 1 and Mode 4 (that is, the present SPH adopted) at instant $t\left(g/H\right)^{1/2}= 10.78$. 
In Mode 1, splashing water particles are "entrapped" by air particles, leading to chaotic interfaces (\frownie), and then causing pressure instability, while Mode 4 maintains uniform particle distribution and smoothed pressure field.
Therefore, strategy $\divideontimes 2$ is the preferred choice for multi-fluid interfaces, especially in violent impact scenarios.
}

\begin{figure}
  \centering
  \includegraphics[width=0.98\linewidth]{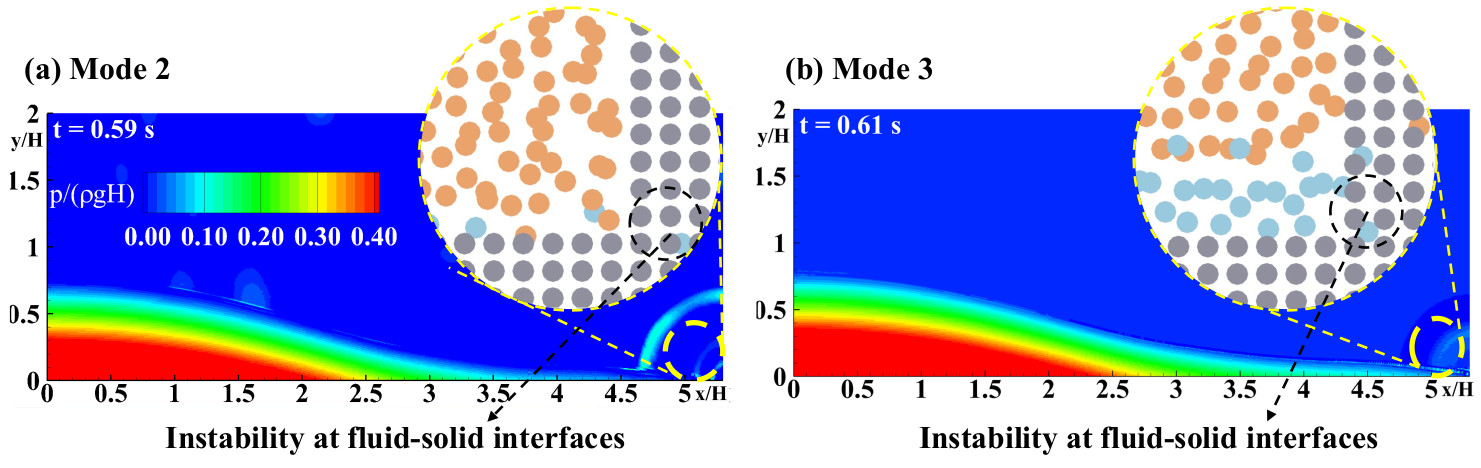}
  \caption{Comparison of pressure fields and particle distribution (enlarged view) at fluid-solid interfaces simulated with Mode 2 and Mode 3.} \label{fig:dambreaking-23-mode} 
\end{figure}

\textbullet{} \textcolor{black}{\textbf{Comparison of two strategies applied at fluid-solid interface}}

\textcolor{black}{
Regarding the treatment of $\delta \bm{u}$-terms at fluid-solid interfaces, Figure \ref{fig:hy-multi-fluid-interface-discussion}, shows that applying strategy $\divideontimes 2$ (i.e., Mode 2 and Mode 3) leads to unstable pressure in the simulation of two-phase hydrostatic simulations.
Furthermore, in the dam-breaking scenario shown in Figure \ref{fig:dambreaking-23-mode}, when the water body with high velocity is about to impact the solid wall, severe instability arises at the fluid-solid interfaces, resulting in unstable pressure fields and particle penetration with Mode 2. 
Similar results are observed simulated with Mode 3, further confirming that the strategy $\divideontimes 2$ shows poor behavior at fluid-solid interfaces (\frownie). Simulations with Mode 2 and Mode 3 fail before water impacting the solid wall. And as discussed in Section \ref{section:results}, using the strategy $\divideontimes 1$ has demonstrated stable behavior at fluid-solid interfaces.
}

\textcolor{black}{
Therefore, based on the above discussion, it is recommended that strategy $\divideontimes 1$ is used for fluid-solid interfaces to ensure stability, while strategy $\divideontimes 2$ is better suited for handling $\delta \bm{u}$-terms in the continuity equation at multi-fluid interfaces.
}

\bibliographystyle{elsarticle-num}

\end{document}